\documentclass[final,12pt]{elsarticle}

\usepackage{booktabs}
\usepackage{cite}
\usepackage{nicefrac}
\usepackage{amsmath,amssymb,amsfonts}
\usepackage{algorithmic}
\usepackage{graphicx}
\usepackage{subcaption}
\usepackage{algorithm,algorithmic}
\usepackage{hyperref}
\usepackage[dvipsnames]{xcolor}

\journal{Biomedical Signal Processing and Control}
\date{20 september 2025}
\begin{document}

\begin{frontmatter}

\title{KDPhys: An Attention Guided 3D to 2D Knowledge Distillation for Real-time Video-Based Physiological Measurement}

\footnotetext{
This paper has been published in \textit{Biomedical Signal Processing and Control}, Elsevier. 
DOI: \href{https://doi.org/10.1016/j.bspc.2025.107797}{https://doi.org/10.1016/j.bspc.2025.107797}.
}

\author[1]{Nicky Nirlipta Sahoo}
\ead{sahoonicky@gmail.com}
\author[1,4]{Sachidanand V S}
\author[1]{Matcha Naga Gayathri}
\author[2,3]{Balamurali Murugesan}
\author[2]{Keerthi Ram}
\author[1,2]{Jayaraj Joseph}
\author[1,2]{Mohanasankar Sivaprakasam}

\affiliation[1]{organization={Indian Institue Of Technology Madras (IITM)},
    country={India}}

\affiliation[2]{organization={Healthcare Technology Innovation Center, IITM},
country={India}}
\affiliation[3]{organization={École de technologie supérieure (ETS), Montreal}, country={Canada}}
\affiliation[4]{organization={Indian Institute Of Science (IISc), Bangalore}, country={India}}

\begin{abstract}
Camera-based physiological monitoring, such as remote photoplethysmography (rPPG), captures subtle changes in the optical properties of the skin due to pulsating variations in blood volume using digital camera sensors. The demand for real-time non-contact physiological measurement has surged, particularly during the SARS-CoV-2 pandemic, to facilitate telehealth and remote health monitoring. Here, we propose an attention-based knowledge distillation (KD) method called KDPhys to extract the rPPG signal from the facial video frames. It effectively distills global temporal information from a 3D convolutional neural network (CNN) based teacher network to a 2D CNN-based student network, utilizing 3D to 2D feature distillation. 
To the best of our knowledge, this is the first implementation of KD in the field of rPPG. Additionally, we introduce a DIstortion Loss including shApe and TimE (DILATE) loss function, which is aware of both shape and temporal information of the rPPG signal. We have conducted qualitative and quantitative experiments on three benchmark datasets. 
Our proposed model significantly reduces complexity, utilizing only half the parameters of existing neural networks while operating 56.67\% faster. With 0.23M parameters, the model demonstrates an overall 18.15\% decrease in Mean Absolute Error (MAE) compared to the current state-of-the-art methods, achieving an average MAE of 1.78 bpm across three datasets at minimal computational cost. Additionally, extensive experiments conducted under diverse environmental conditions and activity types highlight the model's robustness and adaptability.

\end{abstract}


\end{frontmatter}

\section{Introduction}

Physiological measurements of vital signs are integral to daily health monitoring.
The established methods for capturing these signals rely on contact-based sensing techniques like electrocardiography (ECG) and photoplethysmography (PPG) \citep{merino2023heartbeat}. Despite their effectiveness, the requirement for direct skin contact with these sensors can lead to discomfort and inconvenience for patients. As telemedicine continues to evolve, camera-based remote physiological measurements are becoming more suitable for assessment and diagnosis than contact-based sensors \citep{merino2023heartbeat}. This approach relies on capturing subtle color variations in the light reflected from the skin and micro-movements resulting from the cardiovascular pulse generated by heartbeats. Remote photoplethysmography  (rPPG) is a camera-based non-contact physiological measurement method that monitors the change in blood volume by capturing these subtle changes in skin pixel intensity.

\begin{figure}[t!]
\begin{center}
  \includegraphics[width = 1\textwidth]{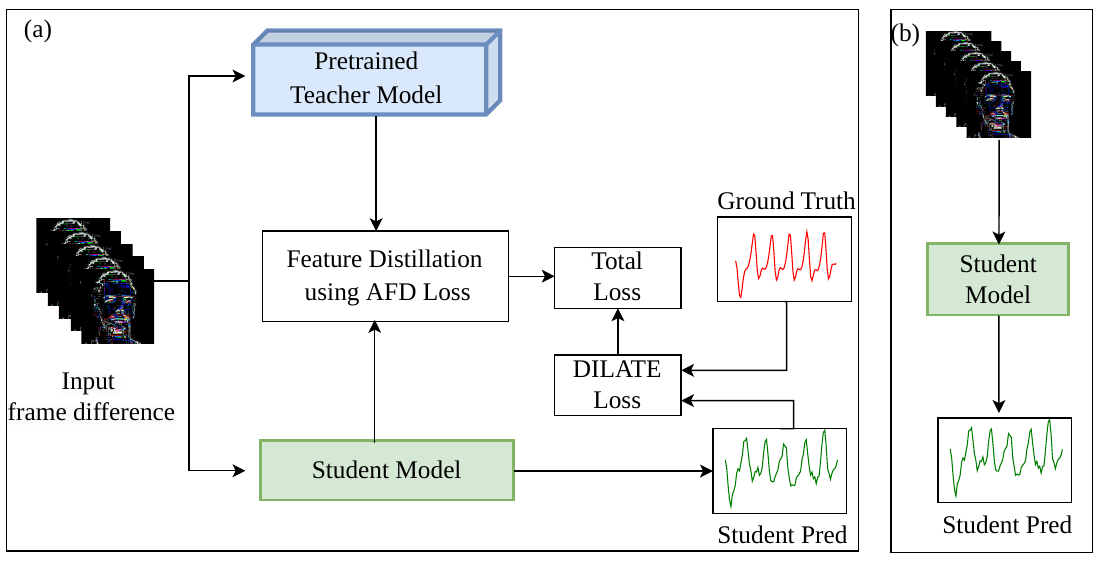}
  \caption{\textbf{Framework of the proposed KDPhys method: (a) Overall proposed training schema with knowledge distillation, (b) Inference process.} The input to both the 3DCNN-based teacher model and 2DCNN-based student model is the frame difference of consecutive frames. The Attention Feature Distillation (AFD) loss facilitates the transfer of feature knowledge from the pretrained teacher model to the student, while the DILATE loss guides the student to align with the ground truth PPG signal. Together, these losses (Total Loss) synergistically enable the student model to capture the significant features of the teacher while adhering closely to the ground truth.}
  \vspace{-2em}
  \label{schema}
\end{center}
\end{figure}

Recent advances in rPPG \citep{tang2023mmpd, benezeth2024video, niu2020video} have introduced various approaches to enhance the sensitivity and reliability of extracting physiological signals from video data, which hold significant potential for applications in cardiology. These methods address key challenges such as motion artifacts, varying lighting conditions, and individual physiological differences. However, practical use of rPPG in cardiology requires further validation in clinical settings, robust performance across populations, and optimized computational efficiency for seamless integration into medical workflows.

Several rPPG methods have been developed in recent years for extracting PPG signals from facial videos. These \textbf{conventional techniques} are primarily based on the principles of PPG extraction in pulse oximeters and can be broadly divided into two categories. The first category comprises blind source separation methods, such as Independent Component Analysis (ICA) \citep{poh2010advancements} and Principal Component Analysis (PCA) \citep{lewandowska2011measuring}, which decompose raw temporal RGB traces into uncorrelated signal sources to extract the pulse signal. The second category includes model-based methods, such as the chrominance-based approach (CHROM) \citep{de2013robust} and plane orthogonal-to-skin (POS) \citep{wang2016algorithmic}, which use color space transformations to extract the blood volume pulse signal.

Various \textbf{end-to-end deep learning} (DL) methods \citep{zhao2023learning, yu2019remoteiccv, huang2021novel, hu2021robust, yu2019remote, sahoo2022deep} have outperformed the conventional hard computing methods in terms of estimating rPPG signal due to several advantages. They eliminate the need for extracting skin patches, thereby simplifying the preprocessing pipeline. Moreover, they effectively address data issues in rPPG estimation, such as noise from facial expressions, eye blinking, and face movements, while accommodating variations in camera parameters and environmental conditions \citep{Lu_2021_CVPR, lu2023neuron}. The robustness of these DL models over conventional methods can be attributed to their ability to decouple the complex relationship between face videos and the rPPG signal.

In real-time applications involving rPPG extraction, 2D Convolutional Neural Networks (\textbf{2DCNNs}) can serve as an effective model \citep{jaiswal2022rppg, vspetlik2018visual} in predicting single-point PPG values in a frame-by-frame manner by extracting local spatial correlations within facial regions. They, however, need additional mechanisms for modeling temporal continuity across frames. 
For example, DeepPhys \citep{chen2018deepphys} adopts a dual-branch attention network based on 2DCNNs. In this, one branch is dedicated to extracting motion information from the difference of consecutive frames (motion branch), while the other branch extracts the facial features using a spatial attention mask (appearance branch). Another example of such a dual branch network is the multi-task temporal shift convolutional attention network (MTTS-CAN) \citep{liu2020multi}, which incorporates a temporal shift module (TSM) \citep{lin2019tsm} in the motion branch. The TSM shifts consecutive input channels of the extracted features along the temporal axis, thus enabling information exchange across multiple consecutive frames. TSM is considered to be parameter efficient and can be used in any 2DCNN model. This capability has been leveraged by EfficientPhys \citep{liu2021efficientphys}, which uses a TSM module to extract the local temporal information. In this, a self-attention module is integrated with the motion branch for single-branch attention-based rPPG extraction.

However, 3D convolutional neural network (\textbf{3DCNN})-based models take a sequence of video frames as input and are able to predict the subsequence of the PPG waveform while incorporating the global temporal information across the frames. Among these, models such as PhysNet \citep{yu2019remote} and CAN3D \citep{liu2020multi} have shown superior performance over 2DCNN models in extracting temporal information with better accuracy while estimating rPPG signals. However, the use of 3D convolution blocks results in higher inference time and computational complexity,  impacting their feasibility for edge deployment.

Given the above benefits and trade-offs in 2DCNN and 3DCNN methods, we put forth the central idea of our method: Can the superior spatiotemporal information learned by a 3DCNN be transferred to a 2DCNN by using a training procedure, offering an alternative that maintains the simplicity of 2DCNN and achieves state-of-the-art performance.

In addressing this, to facilitate knowledge transfer from a more complex 3DCNN model to a simpler 2DCNN model, \textbf{Knowledge Distillation} (KD) emerges as the appropriate approach. Generally, the teacher and student models in KD have similar architectures, with the student model being a simplified or smaller version of the teacher model \citep{murugesan2020kd, hinton2015distilling}. However, recently, with the advent of more KD techniques, there are methods that enable distillation between two completely different teacher-student models by the addition of a suitable adaptation layer, enabling the student to have the feature information of the teacher \citep{liu20213d, liu2022cross}. 
This involves distilling the knowledge learned by the teacher model, including spatiotemporal features, and transferring it to the student model. 

To ensure compatibility between the 3DCNN teacher model and the 2DCNN student model, we incorporated several modifications (Figure \ref{architecture}). The spatiotemporal 3D feature representation is projected onto 2D planes of the student network to align the internal representations of the student model with the teacher model. In the student model, rather than using a fully connected layer to regress the 2D features to 1D PPG signal \citep{liu2020multi, liu2021efficientphys}, here we used the ConvTranspose layer \citep{yu2019remote} along with the TSM and adaptive average pool 2D (AAP) modules. AAP reduces the computational complexity of the model by summarizing feature maps with average values while still capturing the important information for regression tasks. These architectural modifications result in a decrease in the number of parameters in the student model.  
In our implementation, we have integrated attention masks after each layer in both the teacher and student models to extract the significant spatial features. Further, we utilize an attention feature distillation (AFD) \citep{wang2019pay} based KD method, prioritizing important features over traditional distillation methods that assign equal weightage to all features of the teacher model.

In the context of predicting PPG signals, the ability to detect temporal changes is just as crucial as accurately predicting the signal's precise shape. Hence, to deal with non-stationary physiological signals (PPG, ECG), we use DIstortion Loss including shApe and TimE (DILATE) \citep{le2019shape} loss function, which penalizes the shape and the temporal localization errors of change detection \citep{zhang2021dual, le2020probabilistic}.

In summary, our main contributions are as follows:
\begin{enumerate}
 \item We propose the \textbf{KDPhys} framework (Figure\ref{schema}), aiming to elevate the student model's performance by transferring knowledge from the 3D teacher model. To the best of our understanding, this marks the first exploration of distillation techniques for capturing global temporal relationships, ensuring precise rPPG measurement while maintaining the simplicity of the 2D  models.
 \item We employ an attention feature distillation technique that calculates channel weights, facilitating the distillation of important features. Upon distilling these prominent features, the student network exhibits improved accuracy with a decrease in MAE and RMSE by 13\% and 20\%, respectively, compared to the basic KD methods.
 \item We use the DILATE loss function instead of commonly used objectives like Mean Squared Error (MSE) and Pearson loss functions. This choice penalizes both temporal and shape errors in the PPG signal, improving heart rate estimation accuracy with a reduction of MAE by 46\% compared to the Mean Squared error-based loss function.

\item The proposed model was trained and tested on three datasets—UBFC, COHFACE, and PURE—encompassing real-world variations in skin tones, lighting, and activities. It shows strong robustness, achieving a 22.3\% reduction in MAE compared to the state-of-the-art EfficientPhys architecture.

\end{enumerate}

The rest of this article is arranged as follows. Section II provides a brief overview of related work. Section III outlines our model and key principles. Sections IV and V present the experimental results and discussion of the results, respectively. The conclusions are drawn in Section VI.

\section{Related Works}
\subsection{Deep learning for rPPG}
 In 2008, Veruysse \textit{et al.} \citep{verkruysse2008remote} pioneered the extraction of heart rate signals from facial videos, primarily emphasizing the green channel of images captured under ambient light conditions. Since then, various rPPG methods have emerged \citep{wang2024camera, lee2023review, sun2022contrast, xiao2024remote}. These include conventional blind source separation methods \citep{poh2010advancements, lewandowska2011measuring}, model-based methods \citep{de2013robust, wang2016algorithmic}, and deep learning methods \citep{yu2022physformer, liu2021efficientphys, lee2023lstc, casado2023face2ppg, lu2023neuron, wang2022self, wang2023transphys, chen2024actnet, lee2023lstc}. Most previous methods for rPPG prediction typically used DL in the preprocessing pipeline \citep{woyczyk2021adaptive, tang2018non} for tasks such as face detection and ROI tracking. The estimation of rPPG within ROI was then done using conventional methods like ICA \citep{poh2010advancements}, CHROM \citep{de2013robust}. 
 
 \citet{song2021pulsegan} proposed PulseGAN, a GAN framework that refines PPG signals extracted via the CHROM method, producing signals closer to the ground truth reference. However, these methods often relied on extensive preprocessing and postprocessing steps. \citet{yu2019remote} first proposed the use of an end-to-end spatiotemporal network using 3DCNN named PhysNet to extract the rPPG signal from raw facial videos directly. In ETA-rPPGNet \citep{eta2021}, a time domain segment subnet approach was used, segmenting the video into multiple parts before feeding it into a subspace network. Time domain attention was incorporated into the backbone to capture local temporal information. TSCAN+ \citep{li2024ts} enhances the original TSCAN \citep{liu2020multi} by integrating convolutional block attention modules and replacing standard convolution in the appearance branch with depthwise separable convolution, resulting in improved network accuracy. PhysFormer \citep{yu2022physformer} and PhysFormer++ \citep{yu2023physformer++} are video transformer-based architectures designed to enhance rPPG representation by adaptively aggregating local and global spatiotemporal features. They have leveraged the temporal difference transformers and incorporated advanced learning strategies like label distribution and curriculum learning. However, these models are computationally very expensive and, hence, not suitable for real-time deployment. EfficientPhys \citep{liu2021efficientphys} simplifies deployment by utilizing raw frames as input for a 2DCNN-based network. In these 2DCNN networks, a performance gap in heart rate (HR) estimation persisted due to less efficient capture of global temporal information compared to 3DCNN networks \citep{yu2019remote,botina2022rtrppg,zhao2023learning}. 
 The above-mentioned methods face limitations, either in terms of computational efficiency due to the use of 3DCNN-based modules or in capturing global temporal information due to the reliance on 2DCNNs. 

\subsection{Knowledge distillation techniques}

For model compression,  \citet{hinton2015distilling} introduced Knowledge Distillation (KD), demonstrating its efficacy in enabling a small model to attain performance comparable to that of a large model for the same task. This is achieved by training the student model with fewer parameters to emulate a powerful teacher model, minimizing the discrepancy between their soft output values. Their study highlights that soft targets from teacher to student improve generalization compared to conventional hard targets. Subsequently, this idea was expanded to hidden layers, too \citep{romero2014fitnets, komodakis2017paying, passban2021alp}. Since network performance mostly improves with increased depth,  \citet{romero2014fitnets} proposed FitNets, in which intermediate features are used to train student models even if they have a different architecture than the teacher. 
\citet{komodakis2017paying}, and  \citet{murugesan2020kd} employed a regression-based attention feature distillation, minimizing the loss between intermediate features of the teacher and student models.

In prior KD methods \citep{romero2014fitnets, murugesan2020kd}, the student model was typically trained to mimic the features from all pixels with uniform priority. This approach often resulted in the student prioritizing background pixel features, thereby diminishing emphasis on foreground features. This imbalance negatively impacts the overall performance of KD. Various methods have been proposed to address this issue, such as mutual relational knowledge transfer \citep{park2019relational}, structural knowledge transfer using attention feature distillation \citep{komodakis2017paying, ji2021show}, and contrastive learning \citep{tian2019contrastive}. A novel frequency-domain attention mechanism for KD was proposed \citep{pham2024frequency}, enabling global feature alignment between teacher and student models. AFT-KD \citep{yang2023attention} proposes a distillation method that leverages attention and feature blocks to transfer both reasoning process and outcome information, using adaptive loss functions for efficient teacher-student alignment. In our work, we use attention feature distillation (AFD) \citep{wang2019pay}, which not only adjusts the strength of transfer learning regularization but also dynamically determines what are the important features to transfer. This is achieved by assigning weights to individual channels within a feature map based on their utility in the target domain and using these weights to regularize the feature map accordingly.


\begin{figure}[bht!]
\begin{center}
\vspace{2.5pt}
\raggedleft

  \includegraphics[width = 1\textwidth]{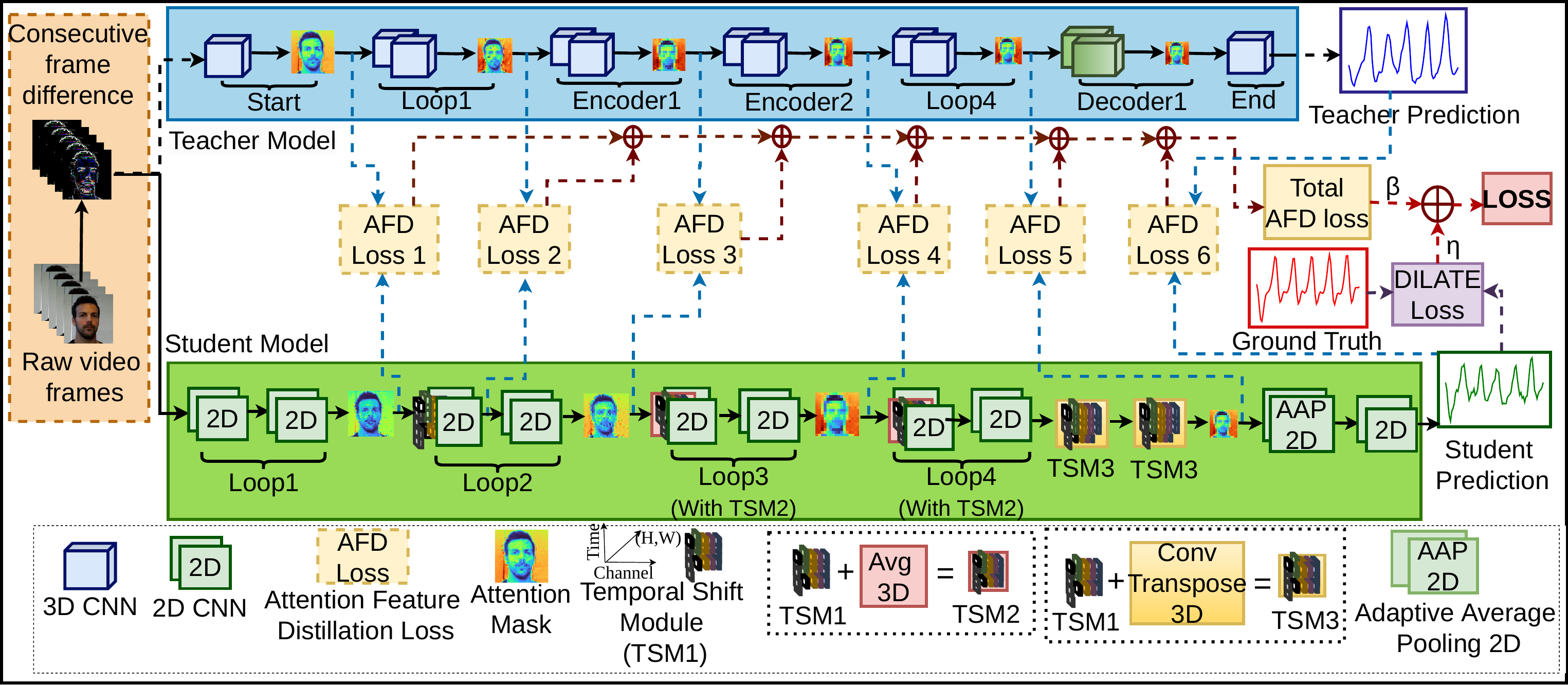} \caption{\textbf{Architecture of our proposed KDPhys network}: First, normalized consecutive frame differences are calculated, which are then used as input to the network. The Upper stream shows the 3D CNN-based teacher attention network, and the lower stream shows the student attention network with the TSM module. The overall architecture shows the AFD-based distillation of features from the teacher to the student network. Along with feature learning, the network is simultaneously trained with respect to the ground truth PPG signal. (The dotted arrows are only included during the training.)}

  \vspace{-2em}
  \label{architecture}
\end{center}
\end{figure}

\section{Methodology}

This section elucidates the comprehensive model architecture of the proposed KDPhys. The overall framework is depicted in Figure\ref{architecture}. The model architecture comprises a teacher model, a student model, and a 3D to 2D Attention Feature Distillation (AFD) based KD technique. First, we discuss the underlying concept behind extracting PPG signals from facial videos, followed by an overview of the architectures of the teacher and student models. Next, we detail the proposed KDPhys method and the loss functions employed for feature distillation. Finally, we conclude the section with the training procedure and the algorithm for the proposed KDPhys.

\subsection{Teacher Model}

In this study, the teacher model is based on the 3DCNN PhysNet model \citep{yu2019remote}. However, we reduced the number of channels in the initial layers from 64 to 32, decreasing the number of parameters of the teacher model. The encoder of the teacher model is further decomposed into two separate sequential encoders (encoder1 and encoder2 in Figure \ref{architecture}) for structural similarity to the student model. This results in efficient distillation. 
Furthermore, we integrated a self-attention module similar to EfficientPhys \citep{liu2021efficientphys} with the teacher model. The self-attention layers are softmax attention layers with 1D convolutions followed by a sigmoid activation function. The normalized self-attention masks from these layers are element-wise multiplied with the output of 3DCNN modules of the Physnet to emphasize the facial regions that are influenced by the changes in the physiological signal. 
We visualize these self-attention masks across all six layers of the network, starting from the shallow layers and gradually moving to the deeper ones. Initially, the masks cover the entire input image, but as we progress deeper, they narrow their focus to specific regions. (Refer Figure S2 of the supplementary for the detail architecture of the teacher model.)

\subsection{Student Model}

The student model builds upon EfficientPhys \citep{liu2021efficientphys} with targeted modifications to enhance its functionality. Notably, our adaptation replaces the fully connected layer in EfficientPhys with a combination of a convolutional transpose (ConvTranspose) layer and a deconvolution layer, similar to the PhysNet model \citep{yu2019remote}. The deconvolution layer integrates a 2D adaptive average pooling (AAP) layer followed by a 2D convolution layer, mirroring the decoder structure in the teacher model. Here, the AAP layer simplifies the network architecture by replacing fully connected layers, thus reducing both complexity and parameter count. 
(Details of this student architecture are presented in Figure S2 in the supplementary material.)

To enable knowledge distillation (KD) from the teacher model, we increased the number of layers in the student model from four to six, modifying the original EfficientPhys architecture. This modification ensures architectural symmetry between the teacher and student models, enabling effective knowledge transfer. 
 Two variants of the Temporal Shift Module (TSM) were introduced in addition to the basic (TSM1) to improve feature alignment: one employing a 3D average pooling layer (TSM2) and the other using a 3D ConvTranspose layer (TSM3). These modules align the feature dimensions of the student model with those of the teacher model, optimizing feature distillation efficiency. Finally, similar to the teacher model, self-attention masks are employed in the student model to emphasize relevant regions, ensuring consistency in feature focus. The specifics of the TSM modules are detailed below:

\begin{figure}[t!]
\begin{center}
\vspace{2.5pt}
\raggedleft

  \vspace{2pt}

  \includegraphics[width = 1\textwidth]{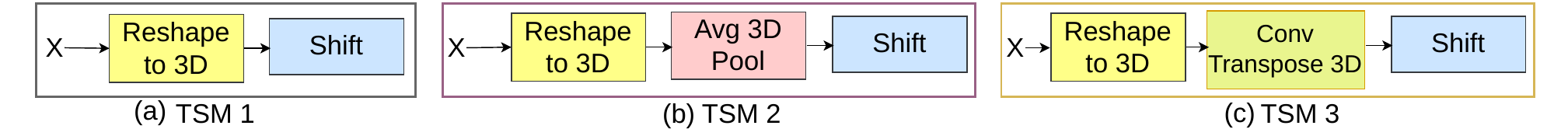}
  \caption{\textbf{Architecture of TSM block variants:} (a) the original TSM block, (b) a variant with integrated 3D average pooling, and (c) another variant incorporating ConvTranspose 3D for architectural symmetry with the teacher model.}

  \vspace{-1.3em}
  \label{tsm}
\end{center}
\end{figure}
\vspace{-2pt}

\paragraph{TSM variants:}
The basic TSM module, as outlined in \citep{lin2019tsm} and illustrated in Figure \ref{tsm}(a) (TSM1), reshapes the input 2D tensor into a 3D tensor by converting the batch size into the depth dimension. The channels of the reshaped tensor are then split into three parts: one part shifts left (advancing by one frame), another shifts right (delaying by one frame), and the third remains unchanged, following the original TSM approach \citep{lin2019tsm}. These three components are then processed through a convolutional layer, allowing a 2D convolutional neural network (CNN) to function as a pseudo-3D CNN without adding extra learnable parameters.

TSM2 (Figure \ref{tsm}(b)) builds upon the basic TSM1 module by incorporating a 3D average pooling layer before the shift operation. This enhancement facilitates temporal pooling, improving feature aggregation and enabling more efficient capture of temporal dependencies.

TSM3 (Figure \ref{tsm}(c)) includes a ConvTranspose3D module with batch normalization, followed by the shift operation to align the student architecture with the teacher model for improved symmetry.

\subsection{KDPhys Framework \& KD Loss Function}

The 3D-to-2D distillation process in the KDPhys method is shown in Figure \ref{architecture}, with the detailed architecture provided in Figure S2 of the supplementary section. Instead of traditional KD techniques \citep{romero2014fitnets}, we used the attention feature distillation (AFD) method \citep{wang2019pay} to improve the transfer of temporal features from the 3DCNNs to the student model. AFD enhances knowledge transfer by using a channel attention mechanism to prioritize important features from the teacher model. This mechanism adaptively regulates the flow of knowledge, ensuring efficient feature transfer with minimal impact on task accuracy and improving the overall effectiveness of the 3D-to-2D distillation process.

Consider a training set $\mathcal{D}$ where each sample $(x,y)$ consists of an input image and the ground truth label. The model is parametrized by $\theta$. The overall loss function for AFD-based KD is defined as,

\begin{equation}
    \mathcal{L}(\theta) = \mathbb{E}_{(x,y)\sim \mathcal{D}}\Big[\big|\big|y-f(x,\theta)\big|\big|_{2}^{2}+R(\theta,x)\Big]
\end{equation}

The regularizer $R(\theta,x)$ is used for feature distillation to apply different penalties for each layer depending on input $x$ and is given by,

\begin{equation}
    R(\theta,x) = \lambda_\text{AFD} \sum_{l\in L'}\sum_{c\in C_{l}} \rho_{l}^{[c]}(x_{l}^{\star}) \big|\big|{(x_{l}^{
\star}-x_{l})^{[c]}}\big|\big|_{2}^{2}
    \label{afd_kd}
\end{equation}

Here, $\lambda_\text{AFD}$ signifies the weightage of the regularizer. $l, c$ represents the current layer \& current channel, and $L', C_l$ denotes the total number of layers \& channels respectively.  $x_{l}^{\star}$ and $ x_{l}$ represent the hint (teacher) and guided (student) layers, respectively. The predictor function $\rho_{l}: \mathbb{R}^{C_l \times H_l \times W_l} \rightarrow \mathbb{R}^{C_l}$ computes the importance of the source activation map for each channel, and assigns weightage accordingly.
\begin{figure}[t!]
\begin{center}
\vspace{-2em}
\raggedleft


  \includegraphics[width = 1\textwidth]{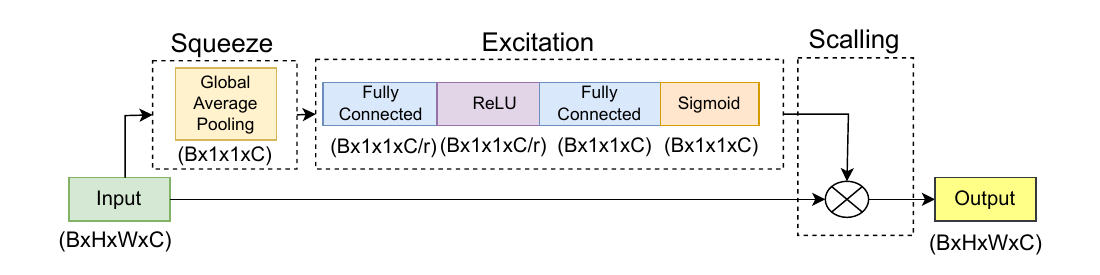}
  \caption{\textbf{Squeeze Excitation module of AFD block}: An input of size ($B \times H \times W \times C$) is fed to the squeeze block, which calculates the spatial average to determine global channel understanding. The excitation block then uses a dense layer followed by ReLU to introduce non-linearity and reduce output channel complexity by a ratio r. This process captures intricate channel dependencies. Finally, weights are applied to the channels by multiplication to compute the output.}

  \vspace{-2em}
  \label{SE_model}
\end{center}
\end{figure}

AFD losses 1 to 6 in Figure \ref{architecture} represent the layer-wise feature losses between the teacher and student models, and the total AFD loss depicts the summation of all these feature losses. The attention $\rho_{l}$ is calculated using squeeze excitation block \citep{hu2018squeeze}, which acts as a content-aware mechanism that re-weights each channel adaptively. Figure \ref{SE_model} shows the architectural unit of the Squeeze Excitation module. In this, the squeeze module
entails the global average pooling of each channel within the feature map, thereby extracting global information. 
The excitation operation computes the channel attention that re-calibrates each channel to enhance the representational capacity of the entire network while mitigating the impact of non-relevant channel information. 

\subsection{Student Loss Function}

In conjunction with weights from the feature-distilled KD model, we employed the DILATE loss function \citep{le2019shape} to train the student model using the ground truth PPG signal. This choice of loss function is motivated by the necessity to capture accurate shape and temporal information for extracting precise physiological parameters from predicted PPG signals.

Loss functions such as MSE and its variants are commonly utilized in training DL networks for extracting rPPG signals \citep{liu2021efficientphys, yu2019remote}. However, these functions may not effectively capture sharp changes in signal characteristics. In contrast, DILATE is specifically designed to address this limitation by reflecting on abrupt changes. It is a differentiable loss function that penalizes shape $(\mathcal{L}_\text{shape})$ and temporal $(\mathcal{L}_\text{temporal})$ localization errors in change detection.
For the predicted output of the model, \begin{math} \hat{y_{i}} = (\hat{y_{i}}^{1},..., \hat{y_{i}}^{k})\end{math}, and corresponding ground truth ${y}_{i} = ({y}_{i}^{1},..., {y}_{i}^{k})$ of length $k$, the DILATE loss function is defined as, 

\begin{equation}
    \mathcal{L}_\text{DILATE} = \alpha \mathcal{L}_\text{shape}(\hat{y_{i}}, {y}_{i}) + (1-\alpha) \mathcal{L}_\text{temporal}(\hat{y_{i}}, {y}_{i})
    \label{dilate}
\end{equation}

 \noindent where the hyperparameter $\alpha \in [0,1]$ is used to have a weighted sum of the spatial and temporal terms. The details about the shape and temporal loss function are outlined in the second section of the supplementary section (Equations S1 and S2 in supplementary).

\begin{algorithm}
\caption{KDPhys knowledge transfer method}\label{alg:kd}
\begin{enumerate}
    \item Step1: Train the teacher network \(f^{T}\) with weights \(\theta^{T}\) using the Mean Squared Error (MSE) loss between the teacher-predicted PPG signal  \(f^{T}(x, \theta^{T})\) and the ground truth PPG signal \(y\):
    \begin{equation*}
    \mathcal{L}_\text{MSE}^{T}\big(y, f^{T}(x, \theta^{T})\big) = \Big|\Big|y - f^{T}(x, \theta^{T})\Big|\Big|_{2}^{2};
    \end{equation*}
    
    \item Step2: Train the student network \(f^{S}\) with weights \(\theta^{S}\) using the DILATE loss and AFD regularization. The overall loss function \(\mathcal{L}_\text{total}\) is defined as the weighted sum of the AFD-based feature loss function and the DILATE loss function with respect to the ground truth:
    
    \begin{equation}
    \mathcal{L}_\text{total} = \beta \times \mathcal{L}_\text{AFD}^{S} + \eta \times \mathcal{L}_\text{DILATE}^{S}
    \label{overall}
    \end{equation} 
    Where, \(\eta\) and \(\beta\) are hyperparameters.
    
    \begin{enumerate}
        \item Distill the features from the teacher layers to the student layers using attention-based feature distillation with AFD regularization \(\mathcal{L}_\text{AFD}^{S}\) as in Eq. \ref{afd_kd}, between intermediate layers to obtain the channel weights based on their significance to the task:  

        \begin{equation}
        \begin{aligned}
        \mathcal{L}_\text{AFD}^{S}\big(f^{S}(x, \theta^{S})\big) = 
        \lambda_\text{AFD} \sum_{l\in L'}\sum_{c\in C_{l}} \rho_{l}^{[c]}\Big(f^{T}(x,\theta^{T})\Big) \times \\
        \bigg|\bigg|\Big(f^{T}(x,\theta^{T}) - f^{S}(x,\theta^{S})\Big)^{[c]}\bigg|\bigg|_{2}^{2}
        \end{aligned}
        \label{afd_final}
        \end{equation}
        
        \item Calculate the DILATE loss \(\big(\mathcal{L}_\text{DILATE}^{S}\big)\) between the student-predicted \(f^{S}(x,\theta^{S})\) and the ground truth PPG signal \(y\), preserving the shape and temporal information between them as defined in Eq. S1 and S2 in the supplementary, respectively:
        \begin{equation}
        \begin{aligned}
        \mathcal{L}_\text{DILATE}^{S} = & \alpha \mathcal{L}_\text{shape}\Big(f^{S}(x,\theta^{S}), y\Big) \\&+ (1-\alpha) \mathcal{L}_\text{temporal}\Big(f^{S}(x,\theta^{S}), y\Big)
        \end{aligned}
        \label{dilate_final}
        \end{equation}
    \end{enumerate}
\end{enumerate}

\end{algorithm}

\subsection{Training Procedure}
This section describes the KDPhys pipeline for transferring knowledge from the teacher to student model (refer algorithm \ref{alg:kd} and Figure S2 in the supplementary). Initially, we trained the teacher model ($f^{T}$) with the MSE loss function and obtained the model parameters $\theta^T$. The pretrained weights from the teacher model are used for distilling features to the student model using attention-based feature distillation (AFD) \citep{wang2019pay} to enable the student network to learn the intermediate feature representation of the teacher. The attention mask from the teacher is obtained using the squeeze and excitation module \citep{hu2018squeeze}. In this work, we have adapted single-step training of the student model ($f^S$) with a total loss function, which is a weighted sum of the AFD loss \ref{afd_final} and the DILATE loss \ref{dilate_final}. Here, the AFD loss function measures the alignment between teacher and student features, and the DILATE loss evaluates the similarity between the student-predicted and ground truth PPG signals.

\section{Experiments}

This section provides a detailed analysis of the proposed KDPhys method, covering experimental requirements, training setup datasets, metrics, and results. We present results from three datasets to validate the model's effectiveness, comparing it against state-of-the-art models. Additionally, we compare computational complexity and latency for real-time analysis and evaluate the proposed KD method and loss function against alternatives, including the effect of feature attention distillation on performance.


\subsection{Experimental Requisites}\
\vspace{-1em}
\subsubsection{Preprocessing}

The raw videos are preprocessed to crop the facial area, ensuring the extraction of maximum physiological pixels. For this, facial landmarks are generated from the first frame using a Haar cascade face detector, and then a larger square region of 160\% width and height of the detected bounding box is cropped (denoted as $c(t)$). For subsequent frames, the multiple-instance learning tracker (MIL Tracker) \citep{babenko2009visual} is used to extract the face region. The difference between consecutive frames is calculated by 
$\nicefrac{c(t+1)-c(t)}{c(t)+c(t+1)+1}$
as in \citep{lin2019tsm}, and normalized by standard deviation. A sequence of such frame differences is used as input for the model. Simultaneously, the derivative of the reference PPG signal is computed and normalized. 
This reference signal is further processed using a Butterworth band-pass filter within a frequency range of 0.5 to 3 Hz before feeding it into the network. The detailed flow diagram of the preprocessing pipeline of the input frames and ground truth PPG signals is shown in Figure S1 of the supplementary section.

\subsubsection{Implementation Details}
The training of the teacher and the student model includes 80 maximum epochs, an Adam optimizer with a learning rate of 0.001, and a batch size of 4. Each mini-batch comprises a sequence of 80 video frames with corresponding label data points. The preprocessed frames are resized to $64 \times 64$ before being given as input to the model. For subject-exclusive cross-validation, each dataset is divided into 50\% of the total subjects for training (22 for UBFC, 81 for COHFACE and 30 for PURE database), 30\% for validation (12 for UBFC, 48 for COHFACE and 17 for PURE database), and the remaining for testing (8 for UBFC, 31 for COHFACE and 12 for PURE database). The UBFC dataset contains subjects with diverse melanin content \ref{ubfc_intra}; hence, the performance improvement on this dataset also signifies its applicability across variations in skin tones. These hyperparameter values are based on recommendations from \citep{le2019shape} and \citep{wang2019pay}, and they have been validated in the ablative study (refer to Section 4.4.9). The stopping criteria for training is determined based on low
validation loss values for extracted PPG signals. The validation curves for the UBFC dataset, trained with different models, are available in Section 3 of the supplementary material. For training and evaluation, the predicted PPG signal is compared with the given ground truth rPPG signal for each database. The model is implemented using PyTorch, and the training process is carried out in a workstation
equipped with an Intel i9 18 core CPU, 192GB RAM, and NVIDIA 24GB GPU.

\subsection{Data Sets}
\begin{figure}[htp!]
\begin{center}
\vspace{-0.5em}
\raggedleft

\includegraphics[width = 1\textwidth]{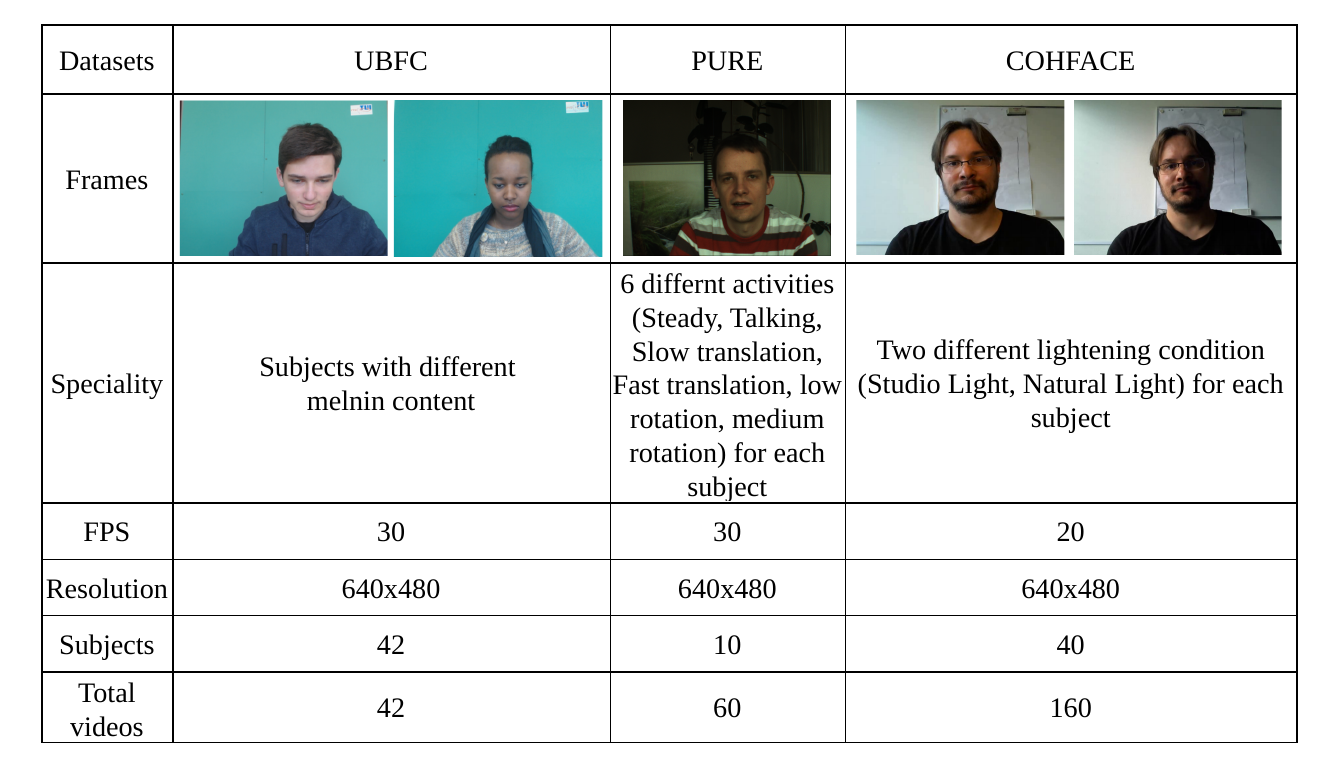}
  \caption{\textbf{Comparison of Differnt datasets:} The datasets utilized in this study (UBFC \citep{bobbia2019unsupervised}, PURE \citep{stricker2014non}, and COHFACE \citep{heusch2017reproducible}) are outlined with some sample frames followed by activity, FPS, resolution, number of subjects, and total number of videos specifications of each dataset.}
  \vspace{-2em}
  \label{dataset}

\end{center}
\end{figure}

In our evaluation, three datasets were utilized to validate the model's performance, and the details are shown in figure \ref{dataset}. The figure shows representative frames of each dataset and their distinguishing characteristics.

\begin{enumerate}
    \item UBFC Database: The UBFC-RPPG database \citep{bobbia2019unsupervised} is a downloadable data set of 42 videos recorded in a realistic environment where subjects were asked to play mathematical games sensitive to time to increase their heart rate. This dataset consists of subjects with different melanin content. These videos were recorded with a low-cost webcam (Logitech C920 HD Pro) at 30 frames per second (FPS) with a resolution of $640 \times 480$ in uncompressed 8-bit RGB format. The ground truth PPG signal and heart rates were collected from a CMS50E transmissive pulse oximeter. 

    \item COHFACE Database: The COHFACE data set \citep{heusch2017reproducible} consists of RGB videos of 40 subjects synchronized with the PPG signal and the breathing rates in two distinct lighting setups. The first setup involved studio lighting, where windows are closed to eliminate natural light, and ample artificial light sources are employed to illuminate the tester's face consistently. The second setup utilized natural lighting conditions. A total of 160 videos were captured, featuring 40 subjects, using a Logitech HD C525 camera at a resolution of $640 \times 480$ pixels, recorded at 20 FPS.

    \item PURE Database: The PURE dataset, introduced by \citet{stricker2014non}, comprises recordings from ten subjects. Each subject underwent six different head motions during recording: steady, talking, slow translation, fast translation, small rotation, and medium rotation. The videos were captured at a frame rate of 30 FPS using an eco274CVGE camera at a resolution of $640\times 480$ pixels. The reference pulse signal (PPG signal), heart rate, and SpO2 were captured using a CMS50E finger clip pulse oximeter at a sampling rate of 60 Hz.

\end{enumerate}
\textit{Ethical Considerations:} The data set utilized comprises publicly available data with proper permissions or data collected under institutional ethical approval. In this study, human facial images were used for experimental purposes. All data collection and processing were conducted in compliance with ethical guidelines, ensuring the protection of the privacy of the participants and their informed consent. No personally identifiable information was used, and all images were anonymized to prevent identification.

\subsection{Metrics}
In order to evaluate the performance of the model, the heart rate (HR) was extracted from the predicted PPG signals. The predicted signals were post-processed using a 1st-order Butterworth bandpass filter with a frequency range of 0.75 to 3 Hz for all the datasets. The HR values were then calculated by employing peak detection in the frequency domain. This involved utilizing the Fast Fourier Transform (FFT) on Hanning windowed signals with a window size of 10 seconds. For the evaluation process, standard metrics were computed, including the Mean Absolute Error (MAE), the Root Mean Square Error (RMSE), and the Pearson correlation (r) \citep{song2020heart} between the calculated HR and the ground truth heart rate ($\text{HR}'$) for an input video of length T. The above metrics can be mathematically defined as follows:

\begin{enumerate}
    \item Mean Absolute Error (MAE):
    \begin{equation}
        \text{HR}_\text{MAE} = \frac{1}{T}\sum_{i=1}^{T}\big|\text{HR}_{i}-\text{HR}_{i}^{'}\big|
    \end{equation}
    
    \item Root Mean Square Error (RMSE):
    \begin{equation}
        \text{HR}_\text{RMSE} = \left({\frac{1}{T}\sum_{i=1}^{T}(\text{HR}_{i}-\text{HR}_{i}^{'})^2}\right)^\frac{1}{2}
    \end{equation}
    \item Pearson Correlation Coefficient (r):
    \begin{equation}
        \text{HR}_\text{r} = \frac{\sum_{i = 1}^{T}(\text{HR}_{i}-\overline{\text{HR}})(\text{HR}_{i}^{'}-\overline{\text{HR}^{'})}}{\sqrt{\sum_{i=1}^{T}(\text{HR}_{i}-\overline{\text{HR}})^2}\sqrt{\sum_{i=1}^{T}(\text{HR}_{i}^{'}-\overline{\text{HR}^{'}})^2}}
    \end{equation}
    Here, $\overline{\text{HR}}$ and $\overline{\text{HR}^{'}}$ represent the mean values of $\text{HR}$ and $\text{HR}'$ over the time period $T$, respectively.

\end{enumerate}

\subsection{Model Evaluations and Performance Analysis}
The results are organized into several sections that provide details on the performance across different datasets, computational costs, and latency evaluations. Additionally, we cover model performance with various loss functions, KD methods, and a comparison of PPG signal quality based on PSNR. Further, an ablative study showcasing results with different hyperparameter values is also included.

\subsubsection{Results on UBFC}

\begin{table}
\centering
\caption{HR estimation results by proposed method and several state-of-the-art methods on UBFC dataset.}
\label{ubfc_intra}
\begin{tabular}{@{}ccccc@{}}
\toprule
\textbf{Method}    & \textbf{MAE ($\downarrow$)}  & \textbf{RMSE ($\downarrow$)}    & \textbf{Pearson ($\uparrow$)} \\ \midrule
CHROM \citep{de2013robust}             & 3.44          & 4.61                       & 0.97             \\
POS \citep{wang2016algorithmic}               & 2.44          & 6.61                       & 0.94             \\
DeepPhys\citep{chen2018deepphys}           & 2.35          & 5.52                   & 0.86             \\
TSCAN \citep{liu2020multi}             & 1.01          & 1.95           & 0.98             \\

PulseGAN \citep{song2021pulsegan}          & 2.09          & 4.42             & 0.97             \\
ETA-rPPGNet \citep{eta2021}        & 1.46          & 3.97              & 0.93  \\         
TSCAN+ \citep{li2024ts} & 0.98 &2.68 &0.97 \\

PhysNet \citep{yu2019remote}           & 1.41          & 3.15            & 0.91             \\

EfficientPhys \citep{liu2021efficientphys}           & 1.00          & 1.75           & 0.98             \\ \midrule
Student (w/o attention)            & 3.03          & 7.1           & 0.79             \\
Student & 0.98          & 1.77           & 0.98             \\
Teacher (w/o attention)           & 1.08          & 2.04           & 0.98             \\
Teacher & \textbf{0.7}           & \textbf{1.49}            & \textbf{0.99}    \\
KDPhys             & \textbf{0.8} & \textbf{1.48} & \textbf{0.99}             \\ \bottomrule
\end{tabular}
\end{table}

We conducted a comparative analysis of our proposed model with conventional methods (CHROM and POS, employed in iPHYS toolbox \citep{iphys}.) and various DL methods (DeepPhys \citep{chen2018deepphys}, TSCAN \citep{liu2020multi}, PhysNet \citep{yu2019remote}, EfficientPhys \citep{liu2021efficientphys}). Specifically, we mention the results for PulseGAN, ETA-rPPGNet, and TSCAN+ based on information from their respective publications \citep{song2021pulsegan, eta2021, li2024ts}. 
The performance metrics are presented in Table \ref{ubfc_intra}. The teacher and KDPhys showed better performance and are highlighted in bold. The table shows that deep learning methods outperformed conventional methods in most cases. 
Here, the student (w/o attention) is modified from EfficientPhys by replacing its fully connected layer with ConvTranspose and deconvolution layers. The results indicate a decrease in performance due to the replacement of the fully connected layer without the inclusion of a self-attention mask, as in EfficientPhys. However, incorporating an attention layer to emphasize important features significantly improves the performance, outperforming the base EfficientPhys model. Further, the results of the teacher (w/o attention) model demonstrate
a notable performance improvement due to the structural modifications made
to the base PhysNet model. Additionally, the integration of the attention
layer, which emphasizes critical features, leads to a significant enhancement
in performance. The results indicate that the proposed student and teacher models outperformed their baseline counterparts, EfficientPhys \citep{liu2021efficientphys} and PhysNet \citep{yu2019remote}, respectively. The incorporation of spatial attention modules in both student and teacher enhances accuracy by focusing on spatial features relevant to physiological signals, facilitating pulse extraction, and reducing background noise. The student model distilled using KDPhys demonstrates substantial improvement, achieving a reduction in MAE and RMSE to 0.8 and 1.48, respectively, compared to the non-distilled student model with corresponding errors of 0.98 and 1.77. This can be attributed to the effective feature distillation using AFD \citep{wang2019pay}, which further reduces MAE and RMSE errors, accompanied by improvements in Pearson correlation. 

Additionally, we have discussed another metric, the Normalized Mean Squared Error (NMSE), a variant of the MSE, in Section 4 of the supplementary material.

\subsubsection{Results on PURE}

\begin{table}[htp!]
\centering
\caption{HR estimation results by proposed method and several state-of-the-art methods on PURE dataset}
\label{pure_intra}
\begin{tabular}{@{}ccccc@{}}
\toprule
\textbf{Method}    & \textbf{MAE ($\downarrow$)}  & \textbf{RMSE ($\downarrow$)}  & \textbf{Pearson ($\uparrow$)} \\ \midrule
CHROM\citep{de2013robust}               & 2.07          & 9.92          & 0.99      \\
POS \citep{wang2016algorithmic}               & 3.14          & 10.57        & 0.95      \\
DeepPhys\citep{chen2018deepphys}           & 4.36          & 6.46         & 0.86          \\
TSCAN \citep{liu2020multi}             & 2.91          & 4.53         & 0.96         \\

PulseGAN \citep{song2021pulsegan}   & 2.28          & 4.29         & 0.99          \\
ETA-rPPGNet \citep{eta2021}         & 2.66          & 6.48         & 0.92          \\
TSCAN+ \citep{li2024ts} & 1.80 &3.45 &0.99 \\  
PhysNet \citep{yu2019remote}           & 2.61          & 4.02          & 0.95            \\
EfficientPhys\citep{liu2021efficientphys}            & 2.07          & 2.61         & 0.98           \\\midrule
Student (w/o attention)            & 4.7           & 6.07         & 0.85              \\
Student & 2.5           & 3.85          & 0.96             \\
Teacher (w/o attention)            & 2.07          & 3.53         & 0.99             \\
Teacher & 1.65          & 3.09         & 0.93             \\
KDPhys             & \textbf{1.61} & \textbf{2.59}  & \textbf{0.99}    \\ \bottomrule
\end{tabular}
\end{table}

We extended our model evaluation to the PURE dataset, which poses unique challenges due to subjects engaging in various tasks.  
The performance comparison of our model with other conventional and state-of-the-art deep learning models is presented in Table \ref{pure_intra}. Here, too, the teacher model has outperformed the baseline PhysNet. The student model demonstrated a slight decrease in performance compared to the original EfficientPhys model, likely due to its simplified architecture, which may present challenges in adapting to diverse tasks. However, the integration of KD results in the addition of knowledge from the 3DCNN network, leading to performance improvements across all baseline models.

\begin{figure}[htp!]
\begin{center}
\vspace{-0.5em}
\raggedleft

\includegraphics[width = 1\textwidth]{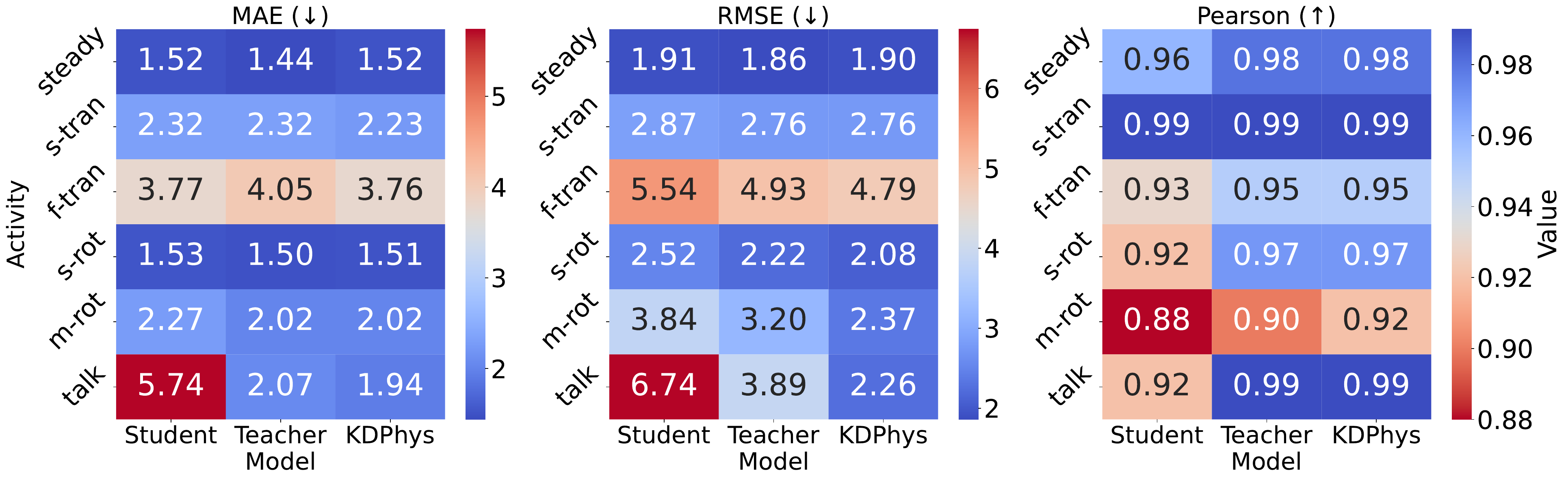}
  \caption{\textbf{Comparison of model performance across six activities—Steady, Talking, Slow Translation (s-tran), Fast Translation (f-tran), Small Rotation (s-rot), and Mid Rotation (m-rot)—is presented using three metrics: MAE, RMSE, and Pearson correlation on the PURE dataset.} The heatmaps visually depict the performance of each model (Student, Teacher, KD) in terms of error and correlation, providing insights into their effectiveness across different activities. In the heatmaps, blue signifies better performance, while red indicates poorer performance.}
  \vspace{-1em}
  \label{heatmap}
\end{center}
\end{figure}

\textbf{Results with respect to different actions in PURE dataset:}
Here, we discuss the performance of the teacher, student, and KD models across various activities for each subject. Figure \ref{heatmap} presents a comparison of the models' performance for different activities, including steady, talking (talk), slow translation (s-trans), fast translation (f-trans), slow rotation (s-rot), and medium rotation (m-rot). 
The following conclusions can be drawn from Figure \ref{heatmap}:
\begin{itemize}
    \item For all metrics in the steady activity, the teacher model outperforms both the student model and KDPhys. However, during fast translation, its performance drops, while the KDPhys model maintains good performance.
    \item In tasks such as talking, the lightweight student model shows limited robustness, as evidenced by a drop in performance. With distillation, KDPhys effectively mitigates this issue, retaining better performance across these scenarios.
    \item Strong linear relationships were observed across most configurations, with Pearson correlation coefficients ranging from 0.95 to 0.99. However, the \textit{`m-rot'} configuration showed a lower Pearson correlation, indicating that the model has difficulty capturing the underlying trends or variability in the data for this specific scenario.
    \item The model demonstrates better performance during steady state, slow translation, talking, and slow rotation, highlighting its robustness and adaptability to these activities.
\end{itemize}

We have also discussed the NMSE metric for the PURE dataset in Section 4, along with a performance comparison of different activities with other state-of-the-art models in Section 5 of the supplementary material.

\subsubsection{Results on COHFACE}
\begin{table}[htp!]
\centering
\caption{HR estimation results by proposed method and several state-of-the-art methods on COHFACE dataset.}
\label{coh_intra}
\begin{tabular}{@{}ccccc@{}}
\toprule
\textbf{Method}    & \textbf{MAE ($\downarrow$)}  & \textbf{RMSE ($\downarrow$)}  & \textbf{Pearson ($\uparrow$)} \\ \midrule
CHROM \citep{de2013robust}               & 7.8          & 12.45         & 0.26             \\
POS \citep{wang2016algorithmic}              & 13.43          & 17.05         & 0.07             \\
DeepPhys \citep{chen2018deepphys}           & 3.91          & 5.59          & 0.62             \\

TSCAN \citep{liu2020multi}             & 4.36          & 6.95         & 0.79             \\
ETA-rPPGNet \citep{eta2021}      & 4.67          & 6.65          & 0.77             \\ 
PhysNet \citep{yu2019remote}           & 3.47          & 5.48                   & 0.78             \\ 
EfficientPhys \citep{liu2021efficientphys}            & 3.34          & 4.92            & 0.65             \\\midrule
Student & 3.55         & 5.74          & 0.74             \\
Teacher & 2.95          & 5.33          & 0.79             \\
KDPhys             & \textbf{2.93} & \textbf{4.82}      & \textbf{0.83}    \\ \bottomrule
\end{tabular}
\end{table}

Similar experiments were conducted on the COHFACE dataset, and the results are presented in Table \ref{coh_intra}. In comparison with the best baseline model, EfficientPhys \citep{liu2021efficientphys}, our model effectively reduced the MAE from 3.34 to 2.93 while maintaining a Pearson correlation of 0.83. The results show that our model has outperformed the current state-of-the-art models across all the metrics.

\begin{figure}[htp!]
\begin{center}
\vspace{-0.5em}
\raggedleft

\includegraphics[width = 1\textwidth]{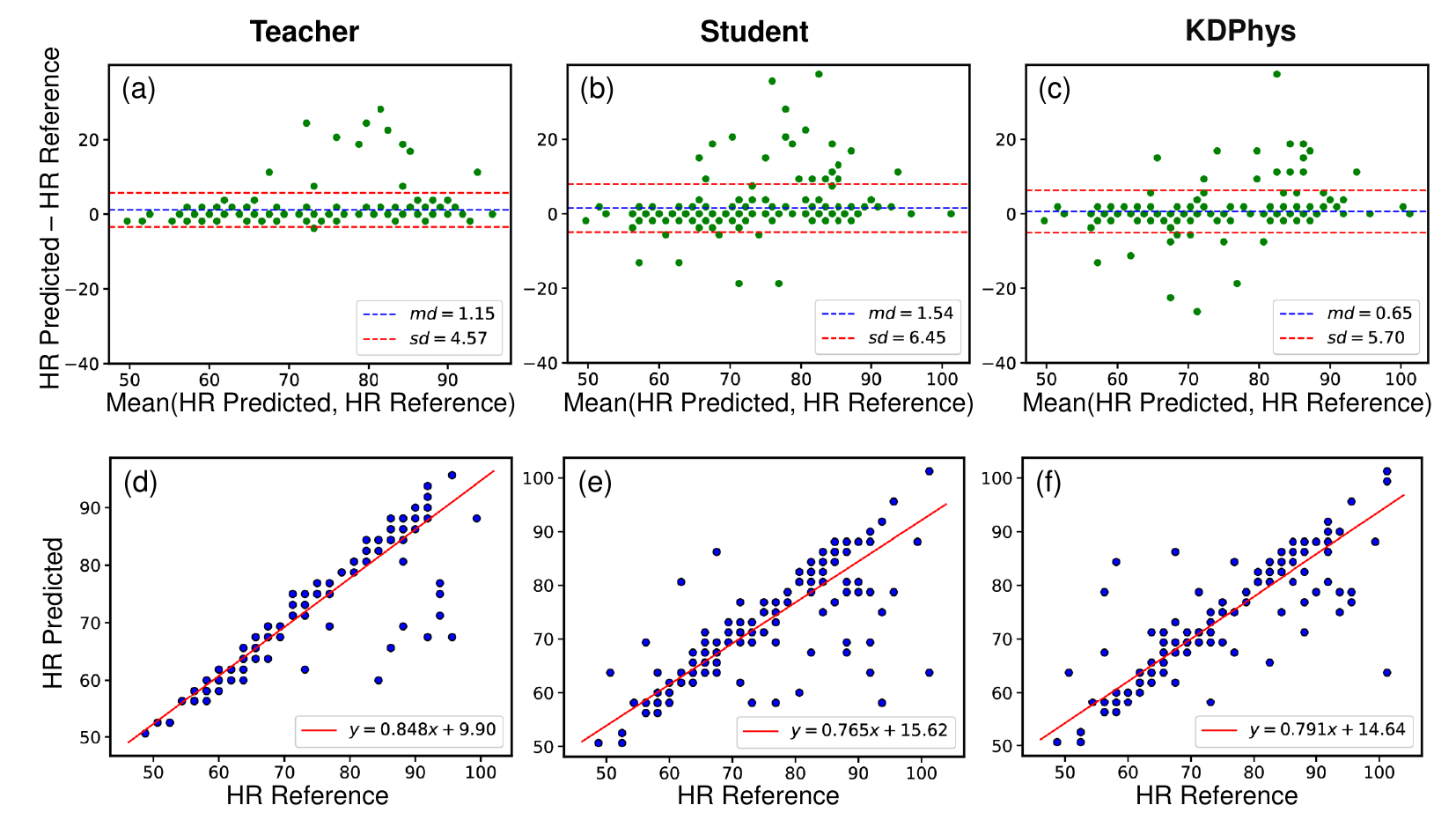}
  \caption{ \textbf{BA and Correlation plot for comparison between predicted and reference heart rate for COHFACE database:} BA and Correlation plot of Teacher model (a,d), Student model (b,e), and KD method (c,f). After using KD, both BA and correlation plots show the performance improvement of the student model with attention-based feature distillation.} 
  \vspace{-2em}
  \label{ba_corr}

\end{center}
\end{figure}
Additionally, we employed statistical plots such as Bland-Altman (BA) plots and correlation plots to better understand the relation between predicted and ground truth HR as illustrated in Figure \ref{ba_corr}. The BA plots of teacher, student, and KD are centralized, with a mean difference (MD) of 1.15, 1.54, and 0.65 bpm, respectively. The standard deviations (SD) fall within an acceptable range of around 5 to 6 bpm. Notably, the BA plot of the student network using KDPhys showed an improvement in the mean difference and standard deviation values, showcasing the efficacy of distillation from the teacher to the student network. The correlation plot further illustrates a robust positive correlation between the predicted and reference HR, with an enhanced slope and minimal bias for the distilled student model compared to the original student model.

\begin{figure}[htp!]
\begin{center}
\vspace{-0.5em}
\raggedleft

\includegraphics[width = 1\textwidth]{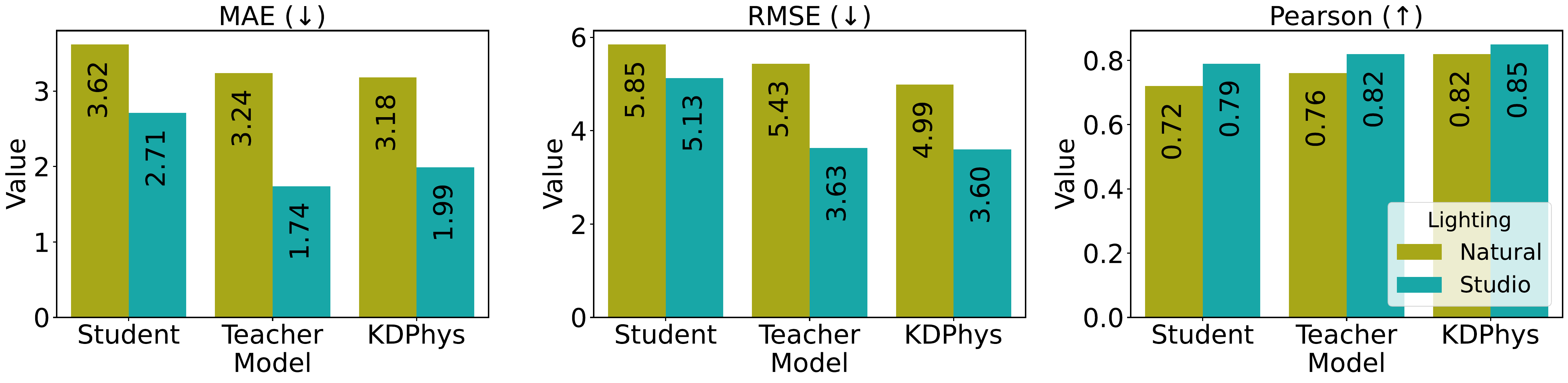}
  \caption{\textbf{Comparison of model performance across different lightening condition using three metrics: MAE, RMSE, and Pearson correlation in the COHFACE dataset.} The barplots provide a visual representation of how each model (Student, Teacher, KD) performs in terms of error and correlation, offering insights into model effectiveness for various activities.}
  \vspace{-2em}
  \label{coh_light}

\end{center}
\end{figure}

\newpage
\textbf{Results of different lighting conditions in the COHFACE data set:}
Here, we have analyzed the performance of the teacher, student, and KDPhys models under various lighting conditions. Figure \ref{coh_light} presents a comparative evaluation of these models based on their performance in different lighting scenarios. The following key observations can be drawn from the above plot:
\begin{itemize} \item The model demonstrates strong performance under both studio and natural lighting conditions.
\item As expected, there is an increase in MAE and RMSE under natural lighting compared to studio lighting. However, this deviation remains within an acceptable range, indicating the model's stability across different lighting conditions.
\item While the teacher model outperforms the student and KDPhys model under studio lighting, KDPhys delivers better results under natural lighting, showcasing its robustness to varying lighting conditions.
\end{itemize}
In addition, we have also discussed the performance comparison of different lighting conditions with other state-of-the-art models in Section 5 of the supplementary material.

\subsubsection{Computational complexity and latency calculation}

\begin{figure}[h!]
\begin{center}
\raggedleft

\includegraphics[trim = 0.5cm 0cm 0cm 0cm,clip,width = 0.95\textwidth]{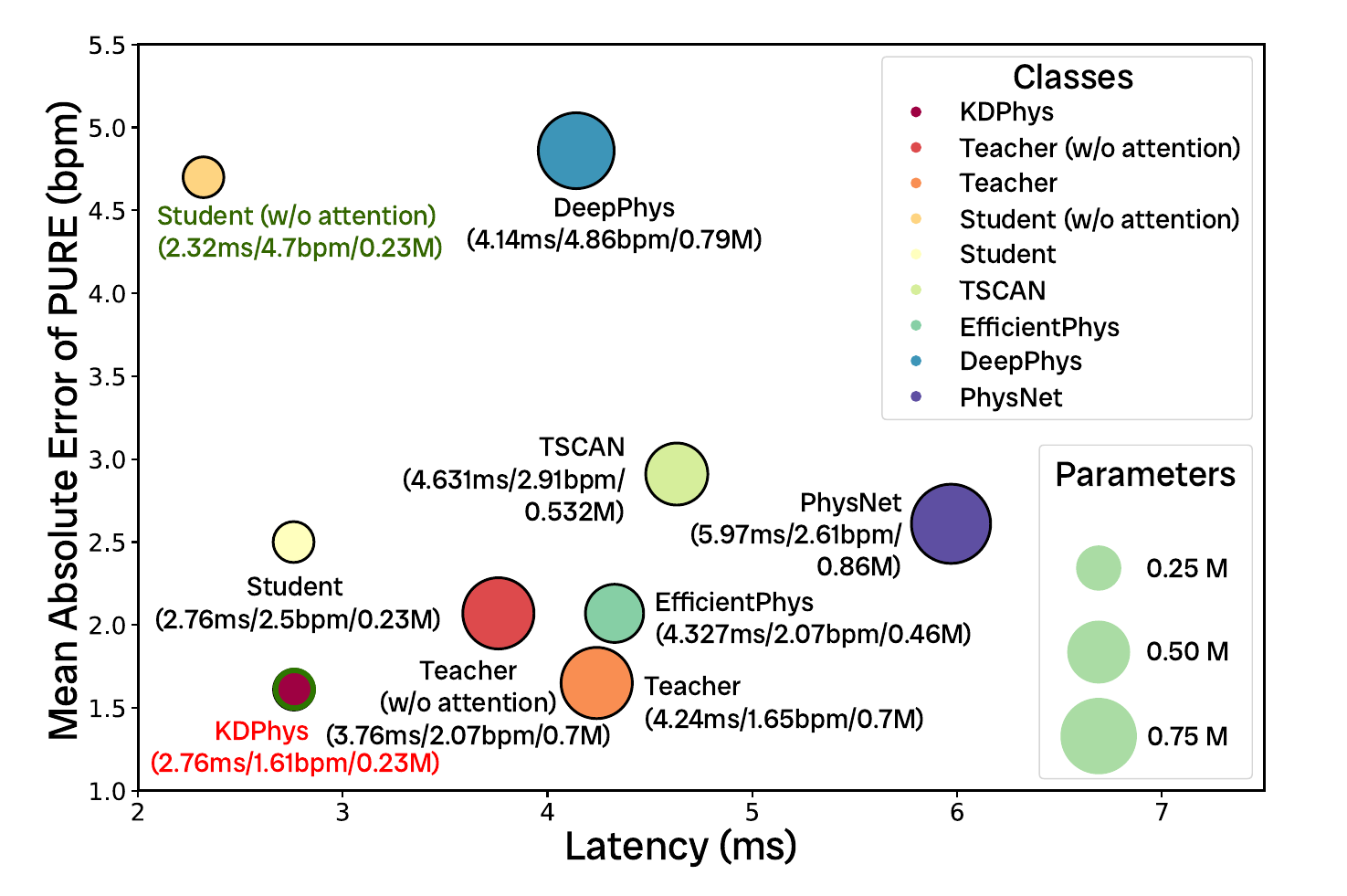}
  \caption{\textbf{Accuracy-Latency Trade-off of nine different methods for PURE dataset.} The X-axis denotes latency in ms, and the Y-axis denotes MAE in bpm. The size of the circle represents the number of parameters in millions (M). Our proposed model (KDPhys) is depicted as a maroon circle with a green circumference, indicating it is the most efficient among the compared models in terms of both accuracy and latency.}
  \vspace{-2em}
  \label{latency}
\end{center}
\end{figure}

Computational complexity and latency are analyzed comprehensively in Figure \ref{latency}, comparing our model with other state-of-the-art DL models. Due to the unavailability of the source code of some models' architectures, direct latency comparison for all architectures is infeasible. However, the models emphasizing computational efficiency and suitability for real-time applications have been taken into consideration. The figure illustrates that the student (w/o attention) model (gold) has half the total number of parameters compared to EfficientPhys. This can be attributed to the use of a deconvolution layer instead of the fully connected layer, which is more computationally complex and prone to overfitting. Secondly, the attention-based KDPhys model (maroon) is approximately 1.5 times faster than the state-of-the-art EfficientPhys model (dark green). Finally, it is observed that the incorporation of attention has not adversely affected the model's complexity and latency; instead, it has significantly enhanced performance by an average of 33.54\% compared to the teacher (w/o attention) and student (w/o attention) model justifying its inclusion in the final model.
Overall, our model surpasses existing state-of-the-art models in complexity and accuracy, proving its suitability for real-time analysis due to its low latency and model complexity.

\begin{figure}[h!]
    \centering
    \includegraphics[width=.75\textwidth]{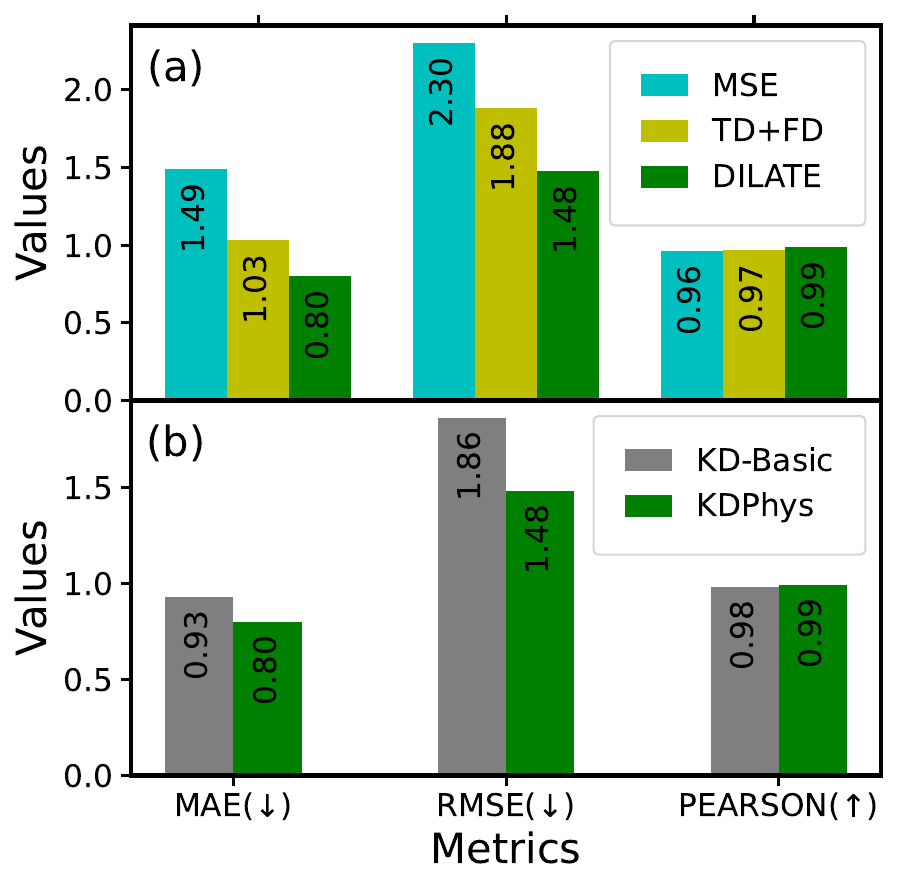}
         
    \caption{\textbf{Plot for comparison between (a) different student loss functions while doing AFD-based KDPhys and (b) basic KD technique and AFD-based KDPhys.} It is evident that the DILATE-based student loss function and AFD-based KD technique (in green) have better performance.}
    \label{fig:5-loss} 
    \vspace{-1em}
\end{figure}

\subsubsection{With different loss functions}

When imputing multiple missing values in a time series, it is essential to ensure that the estimated values closely follow the actual trajectory of the time series.
Therefore, we have utilized the DILATE loss function in our model. 
The bar plot in Figure \ref{fig:5-loss} (a) demonstrates a significant reduction in MAE, achieving an improvement of 46.3\% and 22.3\% compared to the MSE loss function \citep{liu2021efficientphys} and the time- and frequency-domain-based loss function (TD+FD) \citep{Lu_2021_CVPR}, respectively.
The plot demonstrates that utilizing the TD+FD loss function resulted in a 30.87\% decrease in MAE compared to solely employing a loss mainly based on shape function, such as MSE. This can be attributed to the integration of temporal dynamics and the spectral characteristics of the signal while using TD+FD loss. Unlike time-domain loss functions, which may penalize deviations in temporal alignment, DILATE loss allows for some temporal variations while still capturing the essence of the signal's shape. Hence, the DILATE loss function outperforms both of them. This makes it suitable for signals with quasiperiodic characteristics such as rPPG. Moreover, DILATE loss offers adaptability to signal dynamics, enabling the model to adjust the level of distortion allowed in both shape and temporal dimensions based on the complexity of the signal.

\subsubsection{Comparision  of AFD and KD}

We compared the effectiveness of AFD-based KDPhys to the basic KD method \citep{romero2014fitnets}, as shown in figure \ref{fig:5-loss} (b). Here, we have used DILATE as the student loss function for both the KD methods. The results demonstrate that the AFD-based distillation method reduced the MAE and RMSE by 14\% and 20.4\%, respectively. This reduction is attributed to the attention based feature distillation, which emphasizes the key features that contribute most significantly to the regression task.


\begin{figure}[htp!]
    \centering
    \includegraphics[width=1\textwidth]{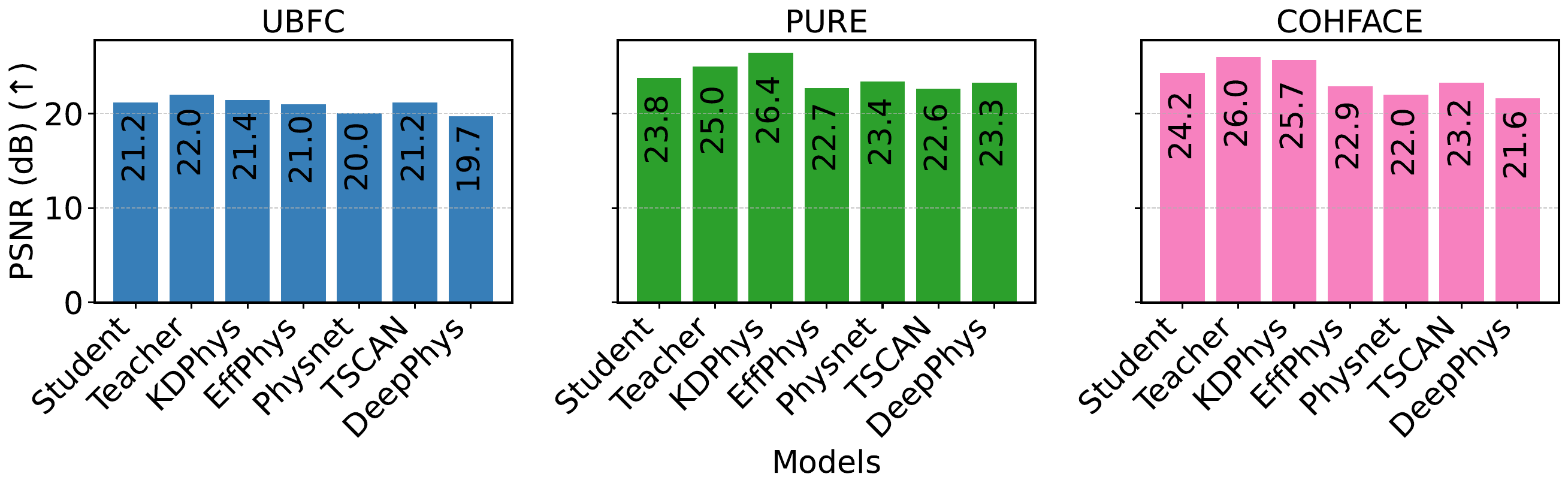}
         
    \caption{\textbf{Plots for PSNR value comparison between different models for (a) UBFC (b) PURE (c) COHFACE datasets.} Here, we have shown the EfficientPhys as EffPhys. The plot shows, the KDPhys model has better rPPG signal quality than all other for challenging datasets like COHFACE and PURE.}
    \label{fig:psnr} 
\end{figure}
\subsubsection{RPPG signal quality comparison with PSNR:}
To evaluate the predicted signal quality, we computed the Peak Signal-to-Noise Ratio (PSNR) values \citep{georgieva2022mathematically} for each model. PSNR measures the ratio of the maximum possible signal power to the power of noise that affects the signal, expressed in decibels (dB). A higher PSNR indicates better signal quality and greater noise resilience. The PSNR is defined as:

\begin{equation}
    PSNR = 10.log_{10}\left(\frac{MAX^{2}}{MSE}\right)
\end{equation}

MAX is the maximum possible signal value, and MSE is the Mean Squared Error between the GT signal and the predicted signal. The PSNR values for the predicted rPPG signals were evaluated across three datasets (UBFC, PURE, and COHFACE) to quantify the fidelity of the signals. Figure \ref{fig:psnr} summarizes the results, showing that our proposed KDPhys and Teacher models outperform existing methods in terms of PSNR. Specifically, the KDPhys model achieves PSNR improvements of 3.7 and 2.8 dB on PURE and COHFACE datasets, respectively, compared to EfficientPhys, the state-of-the-art architecture. In the UBFC data set, the Teacher model achieves the highest PSNR of 22 dB, while the KDPhys model maintains competitive performance. These results demonstrate the robustness of our models, particularly in handling noisy datasets such as PURE and COHFACE, thereby validating their suitability for accurate and reliable rPPG signal reconstruction.

\subsubsection{Qualitative Analysis of rPPG signal}
\begin{figure}[htp!]
\begin{center}
\vspace{-5em}
 \includegraphics[width =1\textwidth]{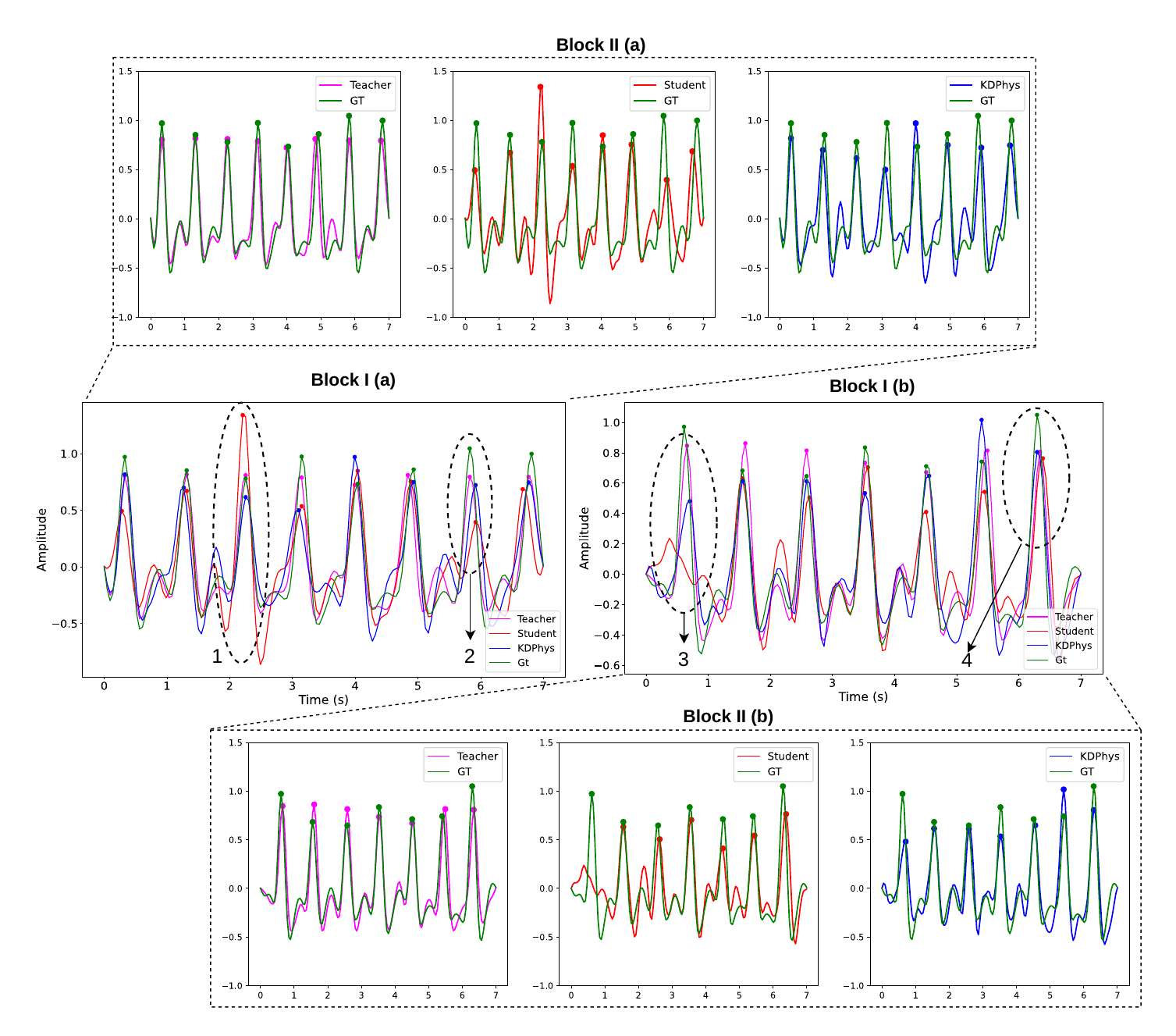}
  \vspace{-1.5em}  \caption{\textbf{Qualitative analysis of PPG waveforms:}  Block I (a) \& (b) present an overall comparison of the extracted PPG signals with GT. The dotted ellipses emphasize the region where performance is different between the student, teacher, and KDPhys method. Block II (a) \& (b) show the individual extracted PPG waveforms of (a) Teacher  (pink), (b) Student (red), and (c) KD (blue) compared to the GT (green). The teacher model shows strong alignment with the GT, and while the student model exhibits some distortions, applying KDPhys significantly improves its alignment.}
  \vspace{-2em}
  \label{ppg_analysis}

\end{center}
\end{figure}

Figure \ref{ppg_analysis} presents the PPG waveform analysis for the teacher model, the student model, and KDPhys compared to the ground truth (GT). For better interpretation, we analyze two PPG subsequences with notable distortions selected from the COHFACE test cases, as shown in Block I(a) and II(a). The heart rate estimation in this analysis is based on the frequency corresponding to the maximum power in the periodogram. Since systolic peaks dominate the signal compared to diastolic peaks, most of the power in the periodogram originates from the systolic peaks. Consequently, our performance analysis focuses primarily on these systolic peaks.

In Block I(a) and  I(b), the dotted ellipses highlight regions where waveform distortions are evident across models. In \textbf{ellipse (1)} of Block I(a), the amplitude of the student model waveform deviates significantly from the GT PPG signal, while both the teacher and KDPhys methods retain the correct amplitude. In \textbf{ellipse (2)} of Block I(a), the predicted PPG amplitudes from the teacher and KDPhys are closer to the GT, with their systolic peaks well aligned. In contrast, the student model exhibits a significantly smaller amplitude, with its peak shifted to the right. The Block II(a) plot shows a clearer picture of the same. Further, in \textbf{ellipse (3)} of Block II(b), the systolic peak of the student model has significant distortion in both amplitude and time in the subsequence compared to the ground truth. The student model's peak is also substantially left-shifted compared to the others, which may reduce the result in heart rate predictions. Similarly, in \textbf{ellipse (4)} of Block II(b), the teacher and KDPhys peaks remain closely aligned with the GT, while the student model peak is shifted to the right.

Block II(a) and II(b) depict a clearer picture, showing that the teacher's signal is better aligned with the GT signal, whereas the student model deviates in various positions. The KDPhys technique has improved the performance of the student model, making it more aligned with the teacher and, hence, with the GT signal.
These observations can be attributed to the following factors:

1. With an input sequence of 80 frames, the 3DCNN model can extract PPG signals that cover at least one period for most rPPG datasets, which are recorded at 60 Hz or lower. Consequently, it can effectively capture global shape and temporal information, making the teacher model more aligned with the GT signal.

2. KDPhys uses global temporal information from the teacher model and local temporal information through the use of the TSM blocks, leading to better temporal alignment compared to the student.

3. The use of the DILATE loss function in KDPhys penalizes shape and temporal information. Hence, it improves the PPG extracted from the student model qualitatively.
\subsubsection{Ablative Study:}
This section analyzes the impact of varying $\beta$ and $\eta$ in the total loss function (Eq. \ref{overall}) and the results for different $\alpha$ values in Eq. \ref{dilate_final}. \\

\begin{table}[!htb]
\centering
\caption{Comparison results based on different $\beta$ and $\eta$ values in the total loss function}
\begin{tabular}{@{}lllll@{}}
\toprule
Hyperparameters & MAE ($\downarrow$)   & RMSE ($\downarrow$) & Pearson ($\uparrow$)\\ \midrule
($\beta, \eta$) = (10,20) & 0.96  & 1.82 & 0.98   \\
($\beta, \eta$) = (5,15)  & 0.938  & 1.78 & 0.99   \\
($\beta, \eta$) = (10,10) & \textbf{0.8}   & \textbf{1.48} &  \textbf{0.99}    \\
($\beta, \eta$) = (15,5)  & 0.93 & 1.78 & 0.99    \\
($\beta, \eta$) = (20,10) & 0.94  & 1.78 & 0.99    \\ 
\bottomrule
\end{tabular}
\label{beta_eta}
\end{table}

\noindent\textbf{Different $\beta$ and $\eta$ in the total loss function:}

The hyperparameter $\beta$ is associated with the AFD-based loss function, designed to enhance intermediate features of the student model and improve the alignment between the predicted rPPG signals of the teacher and student models. Similarly, $\eta$ corresponds to the DILATE loss function, which emphasizes the alignment of the student predicted rPPG signal with the GT PPG signal. Table \ref{beta_eta} summarizes the performance metrics for various combinations of these hyperparameter values. From the table, it can be observed that assigning $\beta$ and $\eta$ as 10 yields best overall performance across different metrics.

\noindent\textbf{Different Alpha values in DILATE loss function:}

\begin{figure}[htp!]
    \centering
    \includegraphics[width=\textwidth]{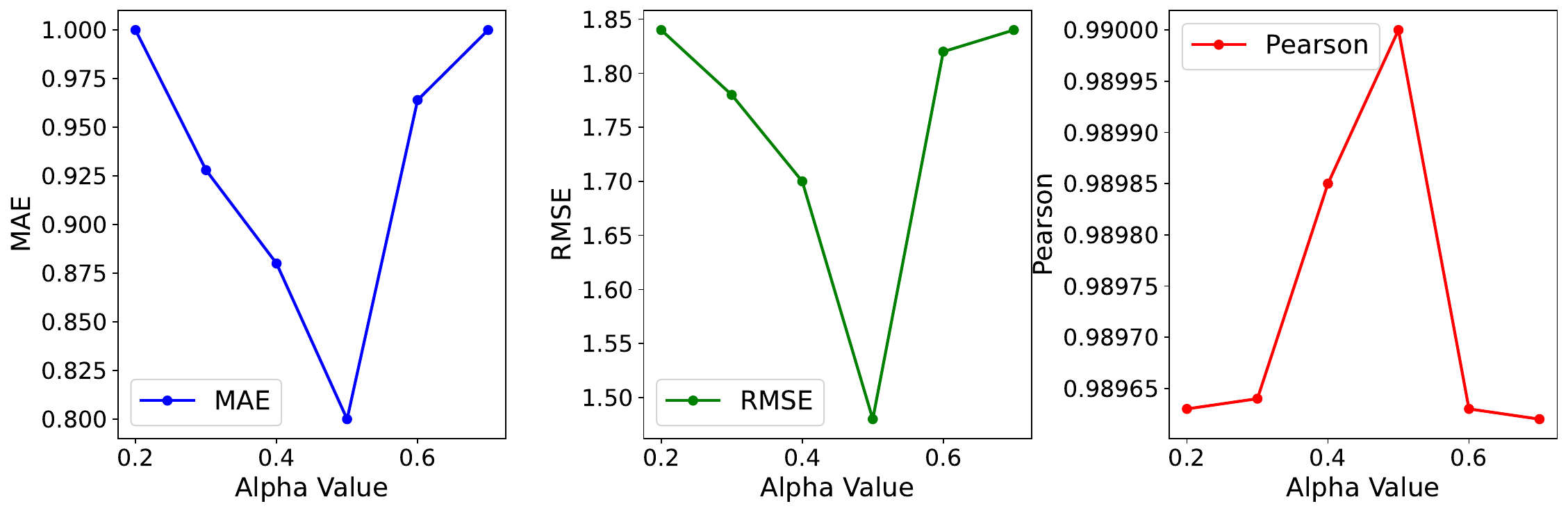}
    \caption{\textbf{Metrics for different values of hyperparameter alpha in the dilate loss function :}This includes MAE, RMSE, and Pearson correlation values for the KDPhys model, evaluated by varying the hyperparameter $\alpha$ in the DILATE loss function.}
    \label{alpha_dilate} 
\end{figure}

We conducted experiments using different $\alpha$ values for the DILATE loss function, where $\alpha$ controls the weight of the shape term, and $(1-\alpha)$ corresponds to the temporal term, as defined in Eq. \ref{dilate}. Figure \ref{alpha_dilate} illustrates the performance metrics for varying $\alpha$ values.

Lower $\alpha$ values emphasize the temporal term, while higher values place greater emphasis on the shape term. The plot indicates that the best result is achieved for $\alpha=0.5$. Further, it is observed that decreasing the weightage of the temporal term ($\alpha > 0.5$) results in more deterioration across performance metrics than increasing it ($\alpha<0.5$).

\section{Discussion}
The results (Table  \ref{ubfc_intra}, \ref{pure_intra} and \ref{coh_intra}) demonstrate that the proposed KDPhys has consistently improved the performance of the student model compared to other state-of-the-art models across all three datasets. From Table \ref{ubfc_intra}, it can be noted that the MAE of the teacher model has reduced by 30\% compared to the baseline EfficientPhys. This can be attributed to the use of 3DCNNs, which extract temporal features from over 80 frames. This span typically covers at least one period of the PPG signal in most rPPG datasets, which are recorded at 60Hz or lower, thereby effectively capturing global temporal information. Additionally, integrating spatial attention emphasizes regions that undergo changes corresponding to physiological variations. The student model effectively learns this information through distillation, achieving performance comparable to the teacher model. Analyzing Table \ref{pure_intra} and \ref{coh_intra}, it is apparent that using a fully connected layer contributes to the better performance of EfficientPhys (average MAE = 2.7) compared to the student model (average MAE = 3.02) for challenging datasets like PURE and COHFACE. 

Nevertheless, this advantage comes at the cost of increased computational demands, as EfficientPhys has twice the total number of parameters compared to the student model, as shown in Figure \ref{latency}. The use of a deconvolution layer instead of the fully connected layer at the output also results in a 46.3\% reduction in latency of the student model compared to EfficientPhys (\ref{latency}). Hence, by replacing the fully connected layer with Adaptive Average Pooling, we significantly reduced the model's parameter count, enhancing its computational efficiency. Additionally, the TSM module in the student model is able to extract temporal information from up to 10 consecutive frames and, hence, captures the local temporal information. In KDPhys, the distillation of features from the teacher network to the student network equips the latter with both global and local temporal information. From Figure \ref{latency}, it can be inferred that KDPhys performs better with a 22.2\% reduction in MAE than the state-of-the-art EfficientPhys model while maintaining lower computational demands, similar to the 2DCNN-based student model with 0.23M parameters.

The attention-based feature distillation (AFD) approach facilitates the transfer of essential features, minimizing the performance gap between the teacher and student. This is demonstrated by an average 23.8\% reduction in the student model's MAE after distillation, as shown in Table \ref{ubfc_intra}, \ref{pure_intra} and \ref{coh_intra}. Incorporating soft labels from the teacher model enables our proposed model to generalize across different subjects within the datasets. The BA plot \ref{ba_corr} illustrates the improvement using the KDPhys technique over the original student model in terms of the mean and standard deviation by reducing it by 57.8\% and 11.62\%, respectively. Also, the correlation plot has showed an improvement in the slope from 0.76 to 0.79 while minimizing the bias from 15.62 to 14.64.

To validate the robustness of the proposed model, we compared its performance across varying lighting conditions (\ref{coh_light}) and different activities (\ref{heatmap}). The UBFC dataset contains subjects with diverse melanin content \ref{ubfc_intra}, and hence, the improvement in performance on this dataset also signifies its robustness to variation in skin tones. The results, as illustrated in the figures (Figure \ref{heatmap}), \ref{coh_light}, demonstrate that the model performs consistently well across lighting scenarios and diverse tasks. Additionally, to assess the stability of the model relative to state-of-the-art models, we conducted a comparative analysis presented in Figures S4 and S5 in the supplementary materials. These comparisons highlight that the proposed KDPhys model exhibits better robustness to environmental variations.

In quasiperiodic signals such as rPPG, when addressing the challenge of imputing multiple missing values within a time series, it becomes crucial that the estimated values not only exhibit reduced average error but also resemble the actual trajectory of the time series. To achieve this, the DILATE loss function is employed instead of conventional MSE-based approaches, as it effectively captures the temporal dynamics of the physiological waveform. The improvements in signal quality achieved through knowledge distillation and the DILATE loss function are evident from the PSNR values of the predicted rPPG signals (Figure \ref{fig:psnr}) and the qualitative analysis (Figure \ref{ppg_analysis}), which show significant improvement of the KDphys compared to the student-predicted PPG signal.

On average, our proposed model demonstrates an 18.15\% reduction in MAE with 0.23M parameters compared to the state-of-the-art model, EfficientPhys, which employs 0.46M parameters. The proposed model, incorporating deconvolution layers, achieves lower latency (2.76 ms)  compared to EfficientPhys, which operates at 4.327 ms. The observed improvements can be attributed to the key factors:

\begin{enumerate}
\item 
The KDPhys framework with attention feature distillation aided the student model in capturing both global and local temporal information across video frames. 
\item The use of a deconvolution layer instead of a fully connected layer has halved the number of parameters of the student model compared to EfficientPhys. Further, with the use of KDPhys, the student model has improved performance (with an average 18.15\% reduction of the MAE) with computational power comparable to the 2DCNN-based student model. 
\item With the use of the DILATE loss function, the student model is able to penalize both shape and temporal distortions, which further helped in improving model performance (46.3\% and 22.3\% reduction in MAE compared to MSE and Temporal and Frequency domain (TD+FD) based loss function, respectively).
\end{enumerate}

 Due to its improved accuracy and notably faster processing speed, our proposed model shows great potential for real-time analysis, positioning it as a valuable asset in telehealth applications and the broader community in computing. Unlike previous hard computing-based conventional methods, our DL model is able to obtain higher accuracy without any complex preprocessing.
 Since the input to the model is difference of the cropped face images, the proposed methods can be deployed for other tasks such as video-based blood pressure measurement, driver monitoring for road safety by emotion and stress detection, sports and fitness monitoring, video-based understanding, and action recognition.
Furthermore, low latency and computational requirements of our model compared to other state-of-the-art models makes it viable for edge deployment. Thus, it can be made accessible to large populations, in low-resource settings, and when in-person consultation with doctors is not feasible. The need for such applications is highlighted during the COVID-19 pandemic. 

The deployment of health-monitoring systems, such as the one described in this study, must carefully consider the implications for individual privacy. Unauthorized use of data to infer sensitive health information, such as heart disease, could lead to ethical and legal challenges. To address this, informed consent must be a cornerstone of any practical implementation. Employees should be fully aware of the data being collected, its purpose, and how it will be protected. Additionally, privacy-preserving techniques, such as edge processing and anonymization, should be employed to minimize risks. These measures align with the principles of fairness and transparency, ensuring that the system is technically sound and ethically responsible.

\section{Conclusion}

We introduced the KDPhys framework, designed to enhance the performance of the 2D student model by distilling knowledge from the 3D teacher model. This exploration of distillation techniques aims to capture global and local temporal relationships, ensuring precise rPPG measurement while preserving the simplicity of 2D models. Employing an attention feature distillation technique facilitated the extraction of crucial features, leading to improved accuracy in the student network compared to the baseline EfficientPhys. Heart rate estimation accuracy was further improved by utilizing the DILATE loss function, which penalizes both temporal and shape distortions in the rPPG signal. Further, KDPhys has shown robustness across different skin tones, lighting conditions, different activities. Through experiments on three diverse datasets—UBFC, COHFACE, and PURE- our proposed model demonstrated a promising average reduction of 18.15\% in error rate while improving the latency by 56.67\% over the existing state-of-the-art EfficentPhys model. 

While the model has outperformed others in terms of metrics and computational efficiency, there is still room for improvement, particularly in handling rapid movements involving fast translations and moderate rotations. Future work could explore advanced post-processing techniques to address motion artifacts in PPG signals. Testing on more diverse populations with more variability and incorporating domain-specific augmentations can further improve generalization. Expanding its application to contexts such as neonatal monitoring, sleep tracking, driver monitoring, stress detection, and emotion recognition presents exciting opportunities for further exploration.

\bibliographystyle{elsarticle-num-names} 
 \bibliography{root}

\end{document}


\begin{frontmatter}

\title{KDPhys: An Attention Guided 3D to 2D Knowledge Distillation for Real-time Video-Based Physiological Measurement}

\author[1]{Nicky Nirlipta Sahoo}
\ead{sahoonicky@gmail.com}
\author[1,4]{VS Sachidanand}
\author[1]{Matcha Naga Gayathri}
\author[2,3]{Balamurali Murugesan}
\author[2]{Keerthi Ram}
\author[1,2]{Jayaraj Joseph}
\author[1,2]{Mohanasankar Sivaprakasam}


\end{frontmatter}
\section{Overall flow diagram and Detailed architecture of KDPhys Model:}
\begin{figure}[htp!]
\begin{center}
  \includegraphics[width = 0.9\textwidth]{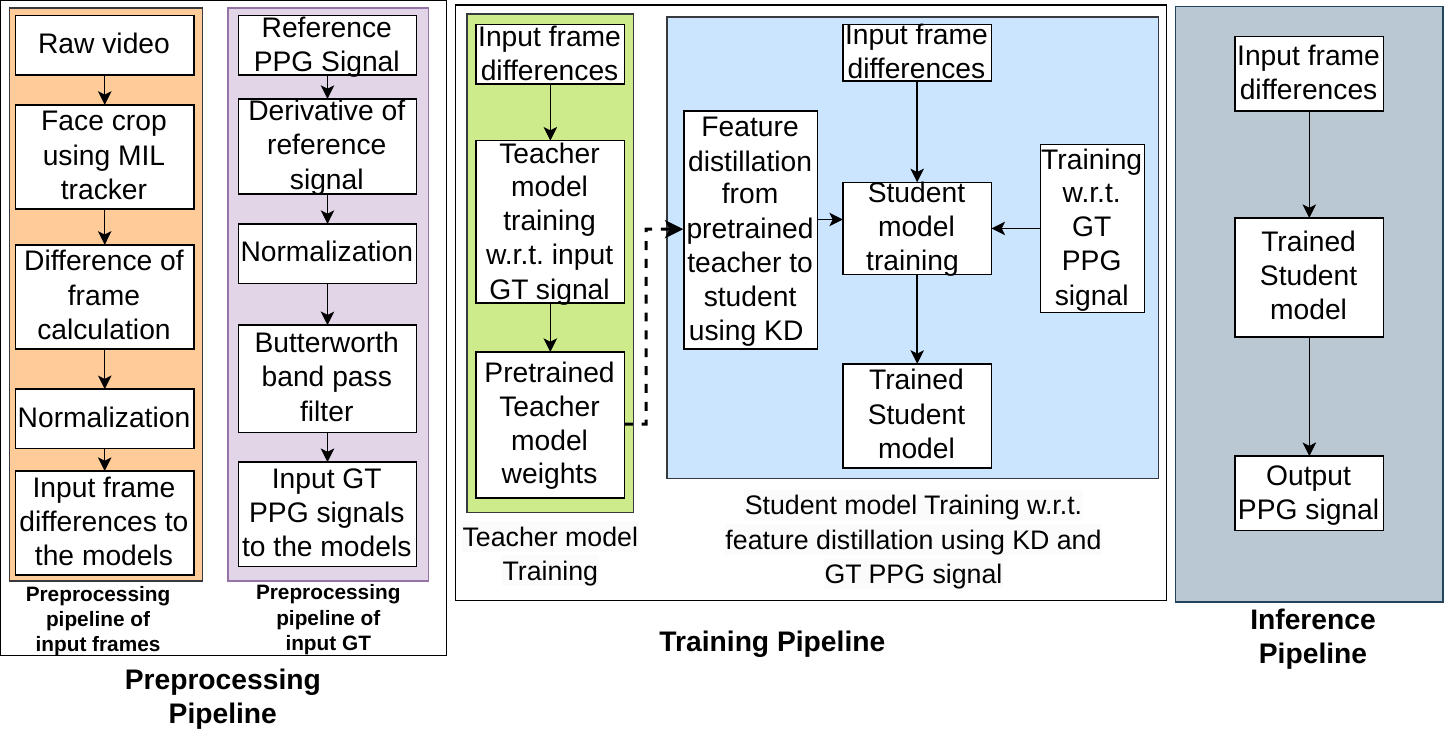}
  \caption{\textbf{The flowchart of KDPhys details three key pipelines:} the \textbf{Preprocessing Pipeline}, which involves preparing input frames and the reference PPG signal for model input; the \textbf{Training Pipeline}, where features extracted from a pretrained teacher model are used to train the student model through knowledge distillation; and the \textbf{Inference Pipeline}, which outlines the process for testing the trained student model on unseen data to evaluate its performance.}
  \label{suppleflowchart}
\end{center}
\end{figure}
\begin{figure}[htp!]
\begin{center}
\vspace{-1em}
  \includegraphics[width = 0.95\textwidth]{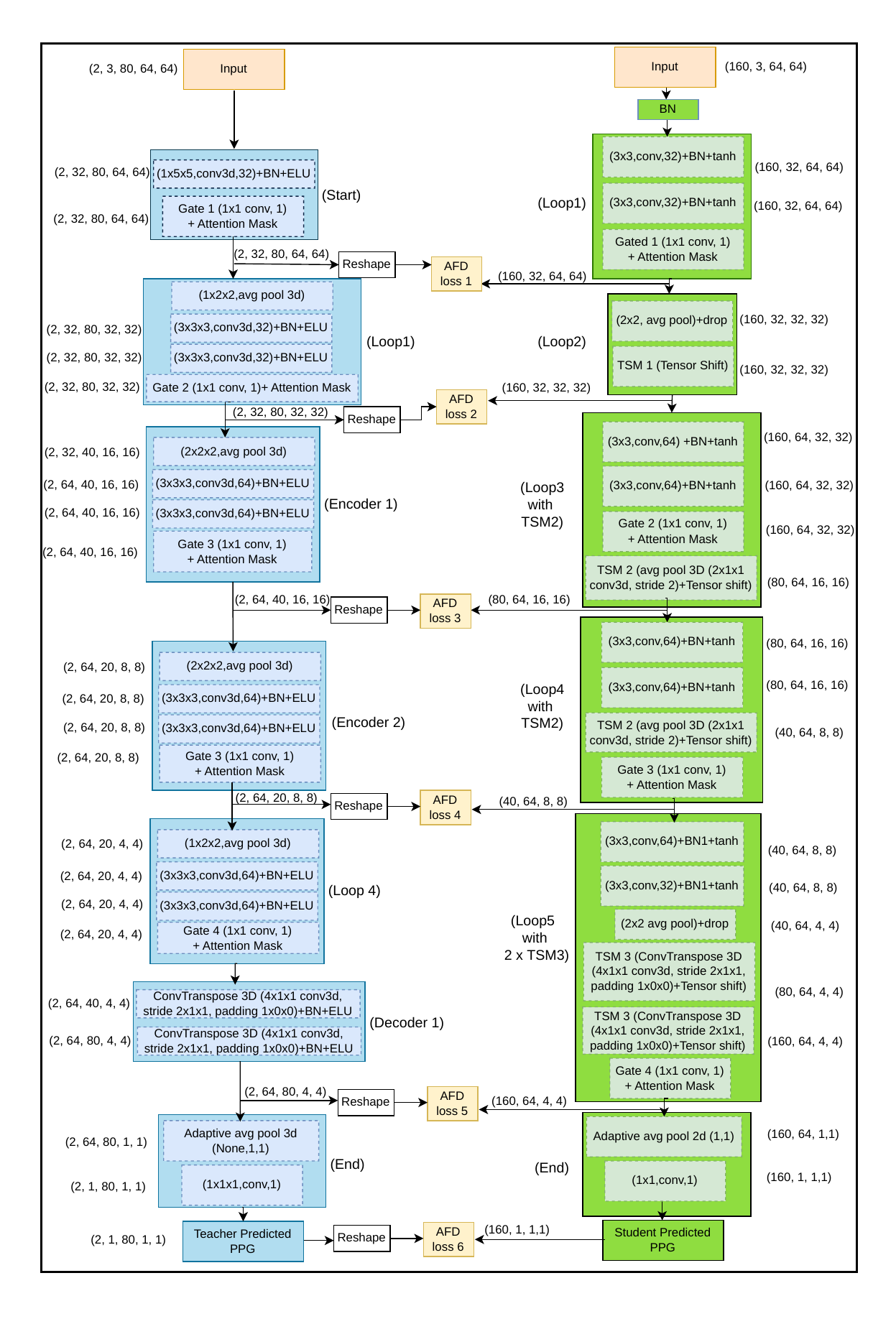}
  \caption{\textbf{Architecture details} of the teacher and student models, along with the knowledge distillation modules.}
  \vspace{-2.5pt}
  \label{suppleschema}
\end{center}
\end{figure}
\newpage
\section{DILATE loss function:}
Here we have detailed about the shape and temporal term of the DILATE loss function. Here the predicted output of the model is considered as \begin{math} \hat{y_{i}} = (\hat{y_{i}}^{1},..., \hat{y_{i}}^{k})\end{math}, and corresponding ground truth ${y}_{i} = ({y}_{i}^{1},..., {y}_{i}^{k})$ of length $k$\\
\noindent\textbf{Shape term:} The shape loss function $\mathcal{L}_\text{shape}$, is based on Dynamic Time Warping (DTW)  \citep{sakore1990dynamic}. DTW mainly focuses on the structural dissimilarity between the predicted $\hat{y_{i}}$ and ground truth ${y}_{i}$, which can be represented by following optimization problem,
 
 \begin{equation*}
     DTW(\hat{y_{i}}, y_{i}) =  \smash{\displaystyle\min_{A \in \emph{A}_{k,k}}}\big\langle A, \triangle\big(\hat{y_{i}}, {y}_{i})\big\rangle
 \end{equation*}
 
 Where the warping path is defined as a binary matrix $A  \subset {{0,1}}^{k \times k}$, with $A_{h,j}=1$ if $\hat{y_{i}}^{h}$ is associated to ${y}_{i}^j$ and 0 otherwise. Pair wise cost matrix is represented as, $\triangle\big(\hat{y_{i}}, {y}_{i}) := [\delta(\hat{y_{i}}^{h}, {y}_{i}^{j})]_{h,j}$, where $\delta$ is the dissimilarity between $\hat{y_{i}}^{h} \text{and }  {y}_{i}^{j}$. $<>$ denotes the inner product between the binary matrix(A) and the pair wise cost matrix. The DTW is made differentiable, applying the smoothed min operator as proposed in \citep{cuturi2017soft}.

 So, the loss term for shape \citep{le2019shape} is defined as,
\begin{equation}
    \mathcal{L}_\text{shape}(\hat{y_{i}}, {y}_{i})
    := -\gamma \log\Bigg(\sum_{A \in \emph{A}_{k,k}}\exp\bigg(-\frac{\big\langle A, \triangle\big(\hat{y_{i}}, {y}_{i})\big\rangle}{\gamma}\bigg)\Bigg)
    \label{Dil_shape}
\end{equation}
Here, the smoothing parameter $\gamma > 0$ is used to make it differentiable.

\noindent\textbf{Temporal term:}
To penalize temporal distortions between the predicted signal $\hat{y_{i}}$ and the corresponding ground truth ${y}_{i}$, the Time Distortion Index (TDI) \citep{frias2017assessing,vallance2017towards} is employed.

The smoothed temporal loss is defined as,
\begin{equation}
\begin{aligned}
&\mathcal{L}_\text{temporal}(\hat{y_{i}}, {y}_{i}) := \frac{1}{Z} \sum_{A\in\emph{A}_{k,k}}\big\langle A,\Omega \big\rangle\exp{\bigg(-\frac{\big\langle A, \triangle(\hat{y_{i}}, {y}_{i})\big\rangle}{\gamma}}\bigg)
\end{aligned}
\label{Dil_temp}
\end{equation}

Where Z is a partition function and is defined as, \\
$Z = \displaystyle\sum_{A\in\emph{A}_{k,k}}\exp{\bigg(-\frac{\big\langle A, \triangle({\hat{y_{i}}, {y}_{i}})\big\rangle}{\gamma}}\bigg)$. $\Omega$ is a square matrix of size $k \times k$ that penalizes each element of $\hat{y_{i}}^{h}$ being associated with a corresponding element $y_{i}^{j}$ when $h \neq j$. The penalization is defined by $\Omega(h,j) = \frac{1}{k^{2}}(h-j)^{2}$, where $k$ is the size of the matrix.
\newpage
\section{Validation curves during training}

\begin{figure}[htp!]
\begin{center}
\vspace{2.5pt}
  \includegraphics[width = 0.85\textwidth]{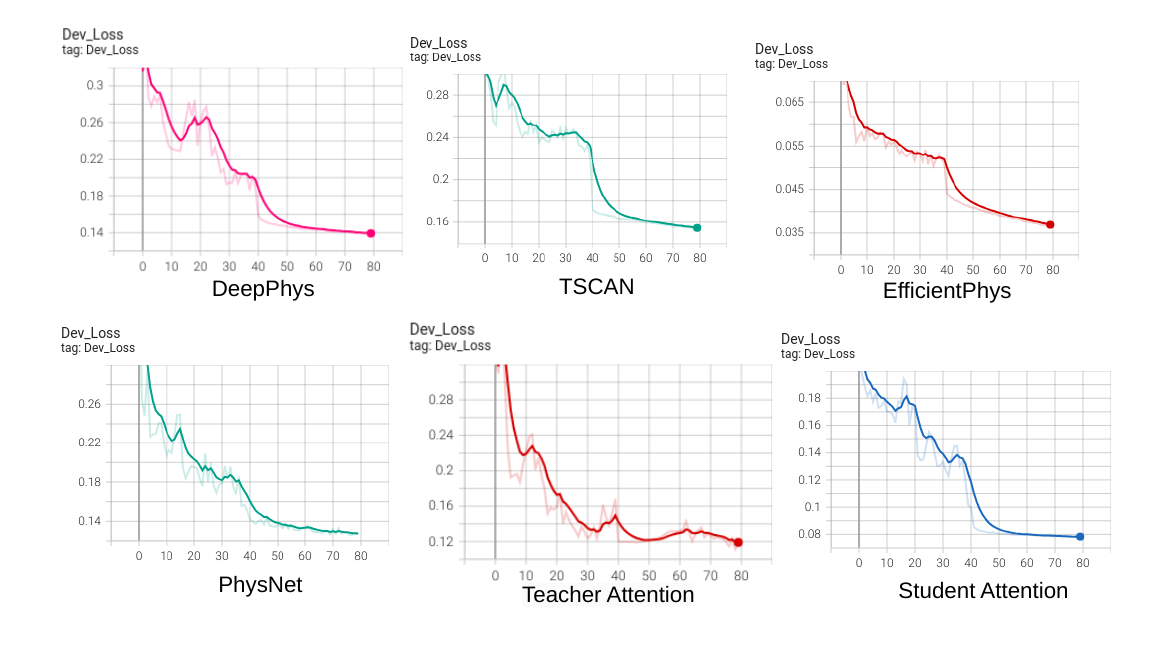}
  \caption{\textbf{Loss curves:} Validation curves for the UBFC dataset trained using different models.}
  \vspace{-2.5pt}
  \label{val_loss}
\end{center}
\end{figure}
\section{Result analysis with NMSE:}
We have calculated the Normalized Mean Squared Error (NMSE) as an additional performance metric to validate the robustness of our proposed model. NMSE provides a normalized error estimate, allowing for effective comparison across datasets and models. This evaluation complements the MAE, RMSE, and Pearson correlation metrics already discussed under section 4.4 in the main text. 

The NMSE between predicted heart rate (HR) and the ground truth heart rate ($\text{HR}'$) is calculated for an input video of length T as follows:

\begin{equation}
    \text{HR}_\text{NMSE} = \frac{{\sum_{i=1}^{T}(\text{HR}_{i}-\text{HR}_{i}^{'})^2}}{{\sum_{i=1}^{T}(\text{HR}_{i}-\overline{\text{HR}^{'}})^2}}
\end{equation}

where,  $\overline{\text{HR}^{'}}$ is the mean of the $\text{HR}'$ values across time T.
Table \ref{nmse_comparison} presents the NMSE values for our model compared to other state-of-the-art models for UBFC and COHFACE datasets. The results highlight that the proposed teacher and KDPhys models achieve significantly lower NMSE values, demonstrating better robustness and accuracy.

\begin{table}[t!]
\centering
\caption{NMSE error metric between estimated HR and the groundtruth HR for the proposed method and several state-of-the-art methods on UBFC and PURE datasets}
\label{nmse_comparison}
\begin{tabular}{@{}ccc@{}}
\toprule
Models                  & \multicolumn{2}{c}{NMSE ($\downarrow$)}      \\ \midrule
                        & UBFC          & PURE          \\
DeepPhys                & 3.61          & 1.21          \\
EffPhys                 & 4.78          & 0.57          \\
PhysNet                 & 1.11          & 0.4           \\
TSCAN                   & 1             & 0.7           \\
Student (w/o attention) & 6.25          & 0.45          \\
Student                 & 2.07          & 0.34          \\
Teacher (w/o attention) & 0.51          & 0.27          \\
Teacher                 & \textbf{0.42} & \textbf{0.25} \\
KDPhys                  & 0.98          & \textbf{0.24}          \\ \bottomrule
\end{tabular}
\end{table}

\begin{figure}[htp!]
\begin{center}
  \includegraphics[width = 1\textwidth]{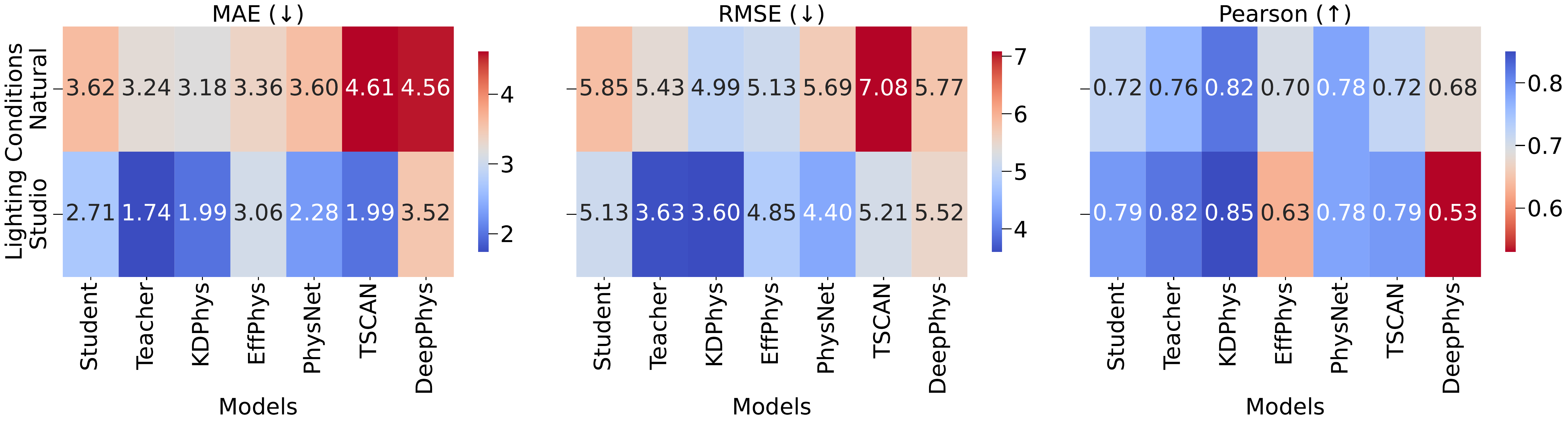}
  \caption{Comparison of model performance across \textbf{different lighting conditions} using three metrics: MAE, RMSE, and Pearson correlation in the COHFACE dataset.}
  \vspace{-5pt}
  \label{heatmap_light}
\end{center}
\end{figure}

\section{Result analysis across existing models in real time environmental conditions:}
\begin{figure}[htp!]
\begin{center}
  \includegraphics[width = 1\textwidth]{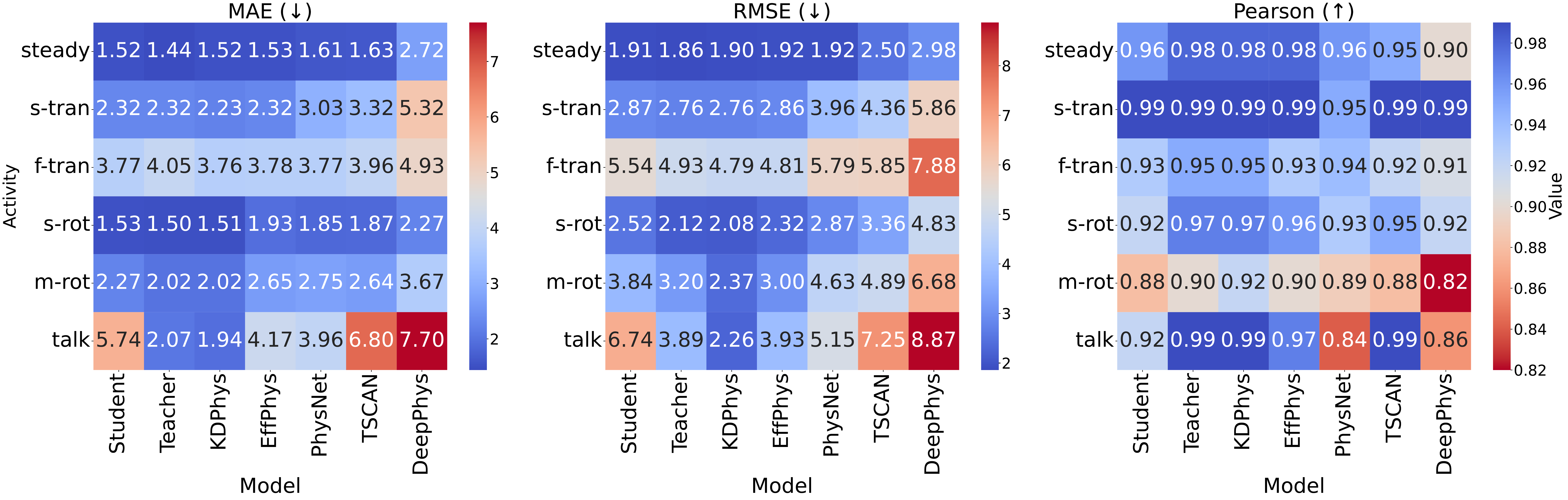}
  \caption{Comparison of model performance across is presented using three metrics: MAE, RMSE, and Pearson correlation on the PURE dataset.}
  \label{heatmap_motion}
\end{center}
\end{figure}

Here, we present two figures: \ref{heatmap_light} and \ref{heatmap_motion}, corresponding to the COHFACE and PURE datasets, respectively, to evaluate the performance of different models under real-world conditions. 

Figure \ref{heatmap_light} highlights the performance gap between studio and natural lighting conditions across various models for the COHFACE dataset. The heatmaps visually depict the performance of each model (Student, Teacher, KD) in terms of error and correlation, providing insights into their effectiveness across different activities. In the heatmaps, blue signifies better performance, while red indicates poorer performance. From this figure, it is evident that KDPhys outperforms all other state-of-the-art models in both cases, showcasing better adaptability to varying illumination.

Similarly, Figure \ref{heatmap_motion} illustrates the results for the PURE dataset under different activity scenarios (Steady, Talking, Slow Translation (s-tran), Fast Translation (f-tran), Small Rotation (s-rot), and Mid Rotation (m-rot)). This figure demonstrates that KDPhys consistently achieves better performance across diverse activities compared to other models.

These findings highlight the robustness and reliability of the KDPhys in handling diverse real-world scenarios, including challenging lighting conditions and different activities.

\bibliographystyle{elsarticle-num-names} 
 \bibliography{root}